\documentclass[aps,prb,twocolumn, superscriptaddress,showpacs]{revtex4-2}

\usepackage{graphicx, amsfonts}
\usepackage{amsmath,amssymb,amsfonts,mathrsfs}
\usepackage{mathtools}
\usepackage{xfrac}
\usepackage{xcolor}
\usepackage{ulem}
\usepackage{soul} 
\usepackage[separate-uncertainty = true,multi-part-units=single]{siunitx}

\begin{document}
\newcommand{\ignorar}[1]{}
\title{Quartic scaling of sound attenuation with frequency in vitreous silica}

\author{Peng-Jui Wang}
\affiliation{Department of Electrical Engineering and Graduate Institute of Photonics and Optoelectronics, National Taiwan University, Taipei 10617, Taiwan}
\author{Agnès Huynh}
\email[Corresponding author: ]{agnes.huynh@insp.jussieu.fr}
\affiliation{Sorbonne Universit\'e, CNRS, Institut des Nanosciences de Paris, F-75005 Paris, France}
\author{T.-C. Hung}
\affiliation{Department of Electrical Engineering and Graduate Institute of Photonics and Optoelectronics, National Taiwan University, Taipei 10617, Taiwan}\author{J.-K. Sheu}
\affiliation{Department of Photonics, National Cheng Kung University, Tainan 70101, Taiwan}
\author{Jinn-Kong Sheu}
\affiliation{Department of Photonics, National Cheng Kung University, Tainan 70101, Taiwan}
\author{ Xavier Lafosse }
\affiliation{Université Paris-Saclay, CNRS, Centre de Nanosciences et de Nanotechnologies (C2N), 10 Boulevard Thomas Gobert, F-91120 Palaiseau, France}
\author{ Aristide Lema\^itre }
\affiliation{Université Paris-Saclay, CNRS, Centre de Nanosciences et de Nanotechnologies (C2N), 10 Boulevard Thomas Gobert, F-91120 Palaiseau, France}
\author{Benoit Rufflé}
\affiliation{Laboratoire Charles Coulomb, Université de Montpellier, CNRS, F-34095 Montpellier, France}
\author{René Vacher}
\affiliation{Laboratoire Charles Coulomb, Université de Montpellier, CNRS, F-34095 Montpellier, France}
\author{Bernard Perrin}
\affiliation{Sorbonne Universit\'e, CNRS, Institut des Nanosciences de Paris, F-75005 Paris, France}
\author{Chi-Kuang Sun}
\email[Corresponding author: ]{sun@ntu.edu.tw}
\affiliation{Department of Electrical Engineering and Graduate Institute of Photonics and Optoelectronics, National Taiwan University, Taipei 10617, Taiwan}
\affiliation{Research Center for Applied Sciences, Academia Sinica, Taipei 115, Taiwan}
\author{Marie Foret}
\email[Corresponding author: ]{marie.foret@umontpellier.fr}
\affiliation{Laboratoire Charles Coulomb, Université de Montpellier, CNRS, F-34095 Montpellier, France}
\date{\today}

\begin{abstract}

Many theories predict a quartic acoustic attenuation increase at sub-THz frequencies in glassy media for the excess vibrational modes known as the boson peak anomaly. Here by introducing phase-sensitive acoustic spectroscopy techniques with a THz bandwidth, we investigate the acoustic properties of vitreous silica at 15 and 300K in the crucial but unexplored sub-THz gap region below the boson peak. Our results indicate a strong negative dispersion starting at \SI{500}{GHz} and the onset of an athermal quartic-frequency-scaling acoustic attenuation term, which emerges above all other thermal losses. 

\end{abstract}


\maketitle


The vibrational dynamics of disordered matter such as glasses exhibit features which differ markedly from those of their crystalline counterparts. In particular, the physics of the damping and dispersion of sound waves in glasses presents a wide variety of phenomena depending on the frequencies $f=\sfrac{\omega}{2\pi}$ and temperatures $T$ at which they are observed. At low $T$, sound is attenuated by the inherent disorder of amorphous solids, while the attenuation is negligeable in defect-free (insulating) crystals. This happens because additional excitations coexist and interact with the sound waves. It can be understood considering the structure of the low-energy portions of glassy potential energy landscapes~\cite{Nishikawa2022}. The tunneling model~\cite{Phillips1972,Anderson1972} postulates that some “entities” have available two nearly degenerate configurations and can tunnel between them. The model explains the anomalous and universal thermal properties observed in glasses well below \SI{1}{K}, including sound absorption. As $T$ rises, a classical regime of thermally activated relaxation (TAR)~\cite{Phillips1987,Gilroy1981,Pohl2002,Damart2018} sets in, resulting in a peak in the acoustic absorption in the 20-200~K range, depending on the glass. TAR processes cover a wide frequency range, from Hz to GHz. At still higher $T$ and $\omega$, the anharmonic interaction of acoustic waves with the thermal phonon bath becomes important~\cite{Rat2005,Vacher2005}.

In addition to these thermal mechanisms, the existence of a Rayleigh-like acoustic phonon scattering regime has long been assumed to be responsible for the existence of the universal plateau~\cite{Zeller1971} in the $T$-dependence of thermal conductivity $\kappa(T)$ of glasses, in the region around \SI{10}{K}. Several theories have been proposed to explain this particular attenuation regime. One of the most important is phonon scattering due to heterogeneities in the elastic constants~\cite{Schirmacher1998,Schirmacher2006,Gelin2016}. The shear elastic heterogeneity has been shown to be dominant~\cite{Leonforte2006}. Recent developments provide evidence that the small scale non-affine displacement fields also play a crucial role in the scattering of acoustic waves~\cite{Caroli2019,Tanguy2002,Szamel2022,Baggioli2022}. An alternative model is resonance with the quasi-localized soft vibrations (QLV) (whose density of states grows as ${\cal D}_\textsc{qlv}(\omega)\propto \omega^4$), which form the boson peak (BP). QLVs are predicted by the soft potential model~\cite{Gurevich2003,Parshin2007} and have recently been demonstrated in numerical studies~\cite{Lerner2016,Shimada2018,Wang2019}. The Rayleigh-type scaling law can also be derived from other physical approaches~\cite{DeGiuli2014,Conyuh2021}, including random matrix models. Numerical studies on glasses of different stability~\cite{Wang2019} verify that, for small phonon wavevectors $q$, a $q^4$-scaling law of sound attenuation is recovered in the zero-$T$ limit for both transverse (TA) and longitudinal (LA) acoustic modes~\cite{Moriel2019}. 
Experimentally, inelastic X-ray scattering (IXS) has made it possible to observe the longitudinal part of the sound waves in some glasses, but only at and above the boson peak frequency $\omega_\textsc{bp}$. This provided direct evidence for the existence of $\Gamma \propto q^4 $ linewidth scaling law over a tiny frequency interval just below $\omega_\textsc{bp}$ in some selected glasses~\cite{Ruffle2003,Ruffle2006,Monaco2009a,Baldi2010}. Instrumental developments have been implemented to enable access to the $q$ range approaching \SI{1}{nm^{-1}}, which is associated with sub-THz frequencies. These include photoacoustic techniques utilising pulse-echo configurations~\cite{Huynh2006,Maznev2012,Devos2008,Klieber2011} and, more recently, involving extreme UV transient gratings~\cite{Foglia2023,Fainozzi2024}. 

In this work, we introduce acoustic time-domain spectroscopy techniques with up to \SI{2.1}{THz} bandwidth to access the acoustic field waveform after transmission through silica glass in the key spectral region corresponding to the low frequency side of BP, i.e. \SI{200}-\SI{900}{GHz}, which is currently inaccessible to experiments. By analysing the acoustic field amplitude, we identify the emergence of an athermal $\omega^4$ scaling law for LA attenuation, which is superimposed on all other thermally induced losses. The dependence of $T$ on the total sound absorption is investigated by measurements at \SI{15}{K} and \SI{300}{K}. The analysis of the acoustic field phase indicates a surprisingly pronounced negative dispersion of the LA mode starting at about \SI{500}{GHz}. Our results highlight the paramount importance of acoustic THz broadband detection techniques with phase sensitivity that measure constant-frequency quantities rather than constant-momentum transfer quantities, while the former better characterizing the glass response as the acoustic wavelength approaches the nanometer scale.


For THz acoustic-transmission-type time-domain spectroscopy techniques, here we generate femtosecond (fs) and picosecond hypersonic pulses and access the ultra-broadband acoustic spectroscopy on v-SiO$_2$ films by either broad-band detection with a piezoelectric single-quantum-well (p-SQW) or by narrow-band high-sensitivity detection with superlattices (SL).

\begin{figure}
	\includegraphics[angle=0,width=0.48\textwidth]{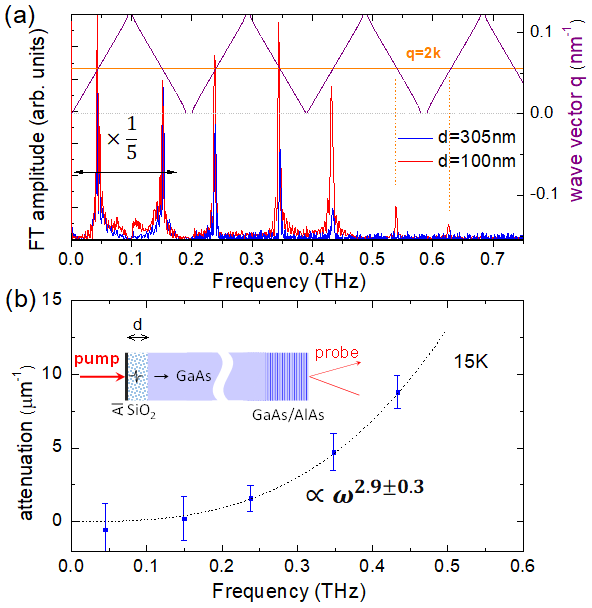}
	\caption{\label{fig:SL} (a) Spectra of the transient changes in reflectivity of the SL, compared to the folded acoustic branches of the SL (black lines), registered at \SI{15}{K} for two SiO$_2$ layers with $d=\SI{100}{nm}$ and $d=\SI{305}{nm}$. The peaks at \SI{0.15}{THz} and \SI{0.05}{THz} have been divided by a factor 5 for clarity.   (b) Attenuation coefficients spectra (Error bar shows the standard error of the mean). The dotted line is a least-square fit.}
\end{figure}

In our first narrow-band high-sensitivity detection scheme, a 30-period \SI{68}{}/\SI{189}{\AA} GaAs/AlAs SL was used to optically detect specific spectral components of the broadband spectrum generated in a 25 nm-thick aluminum film (Fig.~\ref{fig:SL}(b)). Due to the periodicity in the SL, the acoustic phonon dispersion curve is folded into a mini-Brillouin zone whose boundary lies at wavevectors $\pi/d_\textsc{SL}$ ($d_\textsc{SL}$: SL period). The selection rule characterizing the Brillouin frequency in transparent media, $q = 2 k$ ($k$: light wavevector), determines the detection frequencies on the folded branches \cite{Huynh2008,Huynh2015} (Fig.~\ref{fig:SL}(a)). We chose a period $d_\textsc{SL}=\SI{25}{\nano\meter}$ to sample frequencies every $\SI{50}{\giga\hertz}$. Although the spectral content generated in the Al film spans mainly from about \SI{25}-\SI{250}{GHz}, components up to \SI{\sim600}{GHz} can be measured using 780 nm probe light (see details in \cite{Huynh2015}). 
A target v-SiO$_2$ layer was inserted between the GaAs substrate and the Al film, and signal amplitudes for different v-SiO$_2$ thicknesses were compared to access the acoustic attenuation coefficient $\alpha$. v-SiO$_2$ layers were deposited using radio frequency cathodic sputtering with properties close to those of thermal oxide films \cite{Huynh2017}.  Low pump fluences and cryogenic conditions were used to avoid nonlinear acoustic effects and phonon absorption during propagation in the substrate.

Fig.~\ref{fig:SL}(a) shows power spectra registered at \SI{15}{K} for two v-SiO$_2$ layers with thicknesses of \SI{100} and \SI{305}{\nano\meter}. Thicker samples show reduced peak amplitudes, especially at higher frequencies. The attenuation coefficient of detected frequency components shown in~Fig.\ref{fig:SL}(b) are obtained with $\alpha(f)= \frac{2}{d_2-d_1}\ln\frac{A_1(f)}{A_2(f)} $ where $A_i$ and $d_i$ denote the Fourier amplitude and the silica layer thickness of two of the three samples studied (100/305/467\,nm). The overall results have been fitted with an effective power law in frequency using a least squares method. We get $\alpha(f) =(99\pm26) \times f ^{2.9\pm 0.3}$ ($\alpha $ expressed in \si{\per\micro \meter} and $f$ in \si{THz}), represented by the dotted line in~Fig.\ref{fig:SL}(b). It is noteworthy that Dietsche et al. reported a similar result in the $100-250$~GHz frequency range, obtained in a phonon spectroscopy experiment using tunnel junctions~\cite{Dietsche1979}. They found $\alpha(f) = 79 \times f ^{2.9}$ (solid line in Fig.~4 of Ref.~\onlinecite{Dietsche1979}) for (mostly TA) attenuation in silica layers at \SI{1} {K}. The effective exponent of 2.9 is evidence that several attenuation mechanisms with different frequency scaling laws are superimposed in this $f$ and $T$ region.

\begin{figure}[t]
 \includegraphics[angle=0,width=0.47\textwidth]{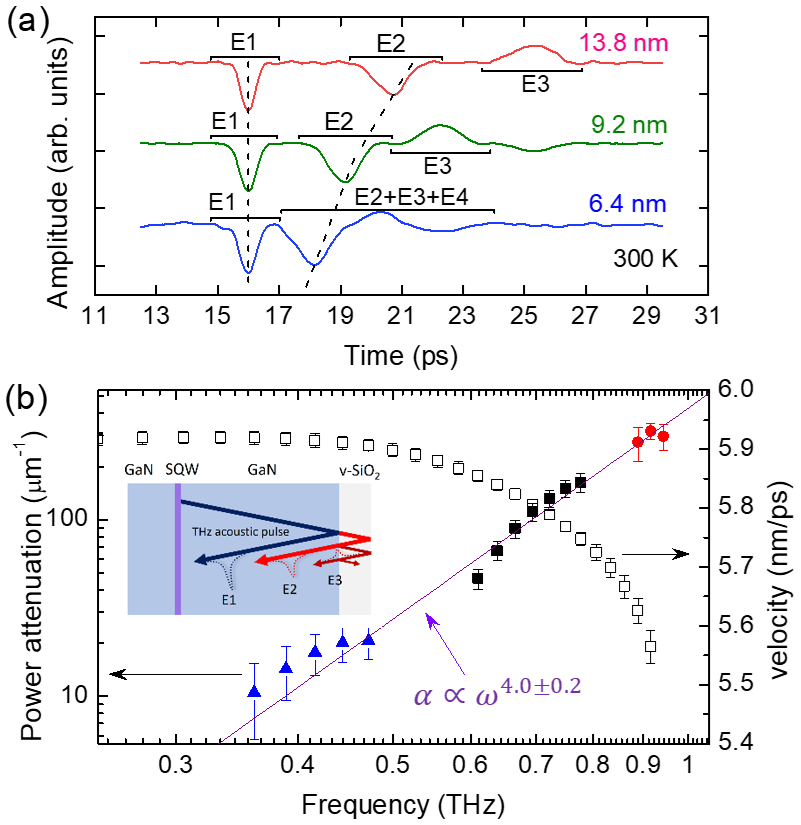}
  \caption{\label{fig:SQW} Inset: Scheme of the cross-section of the GaN/v-SiO$_2$ sample. Successive echoes (E1, E2, ....) resulted from multiple reflections were detected by the p-SQW. (a) Time evolution of echoes after removing background at \SI{300}{K} from samples with film thicknesses of 6.4/9.2/13.8\,nm. (b) Right scale, sound phase velocity  spectra acquired from \SI{9.2}{nm} thickness sample showing a strong negative dispersion above \SI{500}{GHz}. Left scale, sound's power attenuation coefficients of v-SiO$_2$ found at \SI{300}{K} (error bar shows the standard error of the mean). Dataset marked as {\color{red} $\bullet$}, {\color{black} $\displaystyle \blacksquare$}, and {\color{blue} $\displaystyle \blacktriangle$} are calculated from E1/E2 of 6.4nm/9.2nm, E1/E2 of 13.8nm/9.2nm, and E2/E3 of 13.8nm/9.2nm, respectively.The solid line is a least-square fit.}
\end{figure}

In our second broadband detection scheme for filling the sub-THz gap in v-SiO$_2$, we developed an acoustic spectroscopy system analogous to terahertz time-domain spectroscopy (THz-TDS) \cite{Koch2023}. By measuring the time-dependent acoustic field waveform, we captured both the amplitude and phase of the acoustic field. This enables simultaneous measurement of a sample’s acoustic velocity and attenuation spectra across a THz bandwidth by comparing transmission spectra from films of different thicknesses. The novel technique generates and detects fs half-cycle strain pulse using a record-thin \SI{2.5}{nm} InGaN p-SQW embedded under a GaN cap layer, allowing acoustic measurements with a 2.1 THz bandwidth  (see inset Fig.~\ref{fig:SQW}, also see supplementary material \cite{Koch2023,Chichibu1998,Wen2011,Lin2007,Chern2004generation,Klieber2011,Malitson65,Vacher2005,Mante2013thermal}) \cite{Maznev2012,Chou2019,Mante2015,Sun2000,Chern2004,Wen2011,Lin2005}. On the atomically flat surface of GaN, v-SiO$_2$ films with thicknesses $d$ of 6.4, 9.2 and \SI{13.8}{nm} were grown for the acoustic THz-TDS analysis. The interfacial quality and film thickness of the v-SiO$_2$ films were characterized by transmission electron microscopy and atomic force microscopy to ensure well-defined media for acoustic propagation. The half-cycle pulse was partially reflected at GaN/v-SiO$_2$ interface (E1), while other transmitted components undergo multiple reflections inside the v-SiO$_2$, producing successive echoes (E2, E3...)  (see Fig.~\ref{fig:SQW}(a)).

Inside the v-SiO$_2$ film, acoustic waves attenuated by $e^{-\alpha(\omega)\times 2d}$ after a round-trip propagation and lost energy due to interfacial scattering. As frequency increases, scattering losses rise, affecting amplitude but not phase $\phi$. Therefore, E1 and E2 reveal acoustic phase velocity dispersion ($ v=\frac{2d\,\omega }{|\phi_2-\phi_1|}$ ), as shown in Fig.~\ref{fig:SQW}(b), with strong negative dispersion observed above \SI{500}{GHz} upwards.  Furthermore, by cross-comparing samples with different v-SiO$_2$ thicknesses, scattering effects are eliminated under identical interfacial conditions, allowing calculation of the attenuation loss spectra from successive echoes of different silica thicknesses ($d_1$ and $d_2$) by
$
  \alpha=\frac{1}{2(d_2-d_1)} \ln\left( \frac{({E_{i+1}}/{E_{i}})_{d_1}}{({E_{i+1}}/{E_{i}})_{d_2}}\right) $. Thinner v-SiO$_2$ films are suitable for observing higher frequency attenuation spectra but limit low-frequency attenuation due to the need for long propagation lengths for observable attenuation. A trade-off between spectral window, echoes, and thickness must be considered. Intrinsic attenuation data at \SIrange[]{600}{800}{GHz} and around 900 GHz was obtained using E1/E2 values of \SI{13.8}{nm}/\SI{9.2}{nm} and  \SI{6.4}{nm}/\SI{9.2}{nm} samples, respectively. Using E2/E3 of \SI{13.8}{nm}/\SI{9.2}{nm}, attenuation extended down to \SIrange[]{330}{440}{GHz}. Through careful data analysis, three sets of attenuation are obtained (see Fig.~\ref{fig:SQW}(b)). A least-squares fit with an effective power law in frequency gives $\alpha(f) =(0.429\pm 0.033) \times f ^{4.0\pm 0.2}$, $\alpha $ in \si{\per\nano \meter} and $f$ in \si{THz}. Including velocity dispersion ($v(\omega)$), the attenuation can be expressed as a function of $q$. A least-squares fit with an effective power law in $q$ then gives an exponent less than 4: $\alpha(q) \propto q ^{3.7\pm 0.2}$.


\begin{figure}[t]
  \includegraphics[angle=0,width=0.49 \textwidth]{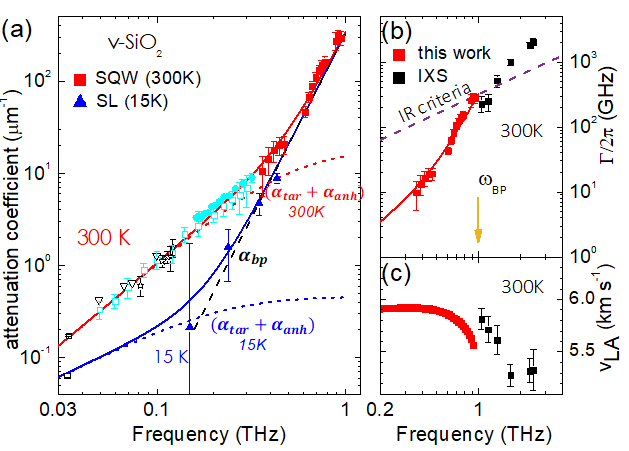}
   \caption{\label{fig:results}
   (a) LA attenuation (of the energy) as a function of frequency at \SI{15}{K} ({\color{blue} $\displaystyle \blacktriangle$}) and \SI{300}{K} ({\color{red} $\displaystyle \blacksquare$}). All other symbols are data from the literature, namely BLS results using visible ({\color{black} $\scriptstyle \square$})~\cite{Vacher2006} and ultra-violet ({\color{black}$\triangle$}{\color{black}$\bigtriangledown$})~\cite{Masciovecchio2004,Benassi2005}, ({\color{black} $\ast$})~\cite{Ruffle2011} incident wavelengths (UV data from Ref.~\onlinecite{Masciovecchio2006} are excluded as unexplained~\cite{Ruffle2011}). The symbols ({\color{cyan} $\bullet$})~\cite{Ayrinhac2011a} and ({\color{cyan} $\scriptstyle \square$})~\cite{Klieber2011} are previous data from picosecond acoustics. The lines are explained in the text. (b) Spectral linewidth $\Gamma =\alpha v$ and (c) velocity $v$ of quasi-LA phonons measured by IXS at \SI{300}{K}~\cite{Baldi2011} ($\displaystyle \blacksquare$) compared to these results ({\color{red} $\displaystyle \blacksquare$}). The dashed line in (b) represents the Ioffe-Regel (IR) limit where the phonon mean free path reaches half the phonon wavelength, i.e. for $\Gamma \cong \sfrac{q v}{\pi}$.}  
\end{figure}

The overall new results, shown as solid symbols in Fig.~\ref{fig:results}(a), extend the existing experimental data on sound attenuation in silica glass up to near THz frequencies. As a reminder, below a few hundred GHz, $\alpha(f,T)$ is known to be driven by two main dissipation mechanisms : TAR of structural ``entities" and anharmonic interactions. TAR processes  are known to dominate the damping in the sonic and ultrasonic regimes. They produce an attenuation $\alpha_\textsc{tar} \propto \omega^2$ for $\omega \tau_\textsc{tar} \ll 1$ and $\alpha_\textsc{tar} \propto \omega^0$ for $\omega \tau_\textsc{tar} \gg 1$, where $\tau_\textsc{tar}$ is the characteristic time of the relaxing entity. A large distribution of relaxing entities is in fact assumed thus leading to an effective power law $\alpha_\textsc{tar} \propto \omega^{n}$ whose exponent $n$ varies with $T$ and satisfies $n(T) \leq 2$. The anharmonic losses result from the interaction of sound with the thermally excited vibrations in the medium at $T$. Following the Akhiezer treatment, the interaction can also be treated as a relaxation process characterised by the mean thermal lifetime $\tau_{th}(T)$ of the dominant vibrations~\cite{Vacher2005}. At room $T$, $\tau_{th}$ is very short so that $\omega \tau_{th} \ll 1$ for frequencies up to at least one hundred GHz, leading to $\alpha_{anh} \propto \omega^2$. The power exponent is nearly independent of $T$ as long as $\omega \tau_{th} \ll 1$. An estimate of the relative contributions of the TAR and anharmonicity processes to the sound absorption in v-SiO$_2$ was given in Ref.~\onlinecite{Vacher2005}. The dotted lines in Fig.~\ref{fig:results}(a) are estimates of the overall contribution $(\alpha_\textsc{tar} + \alpha_{anh})$ for \SI{300}{K} and \SI{15}{K} using the models and the parameters described in ~\cite{Vacher2005,Ayrinhac2011a,Huynh2017}. The new data therefore highlight the presence of the additional dissipation process,  $\alpha_{bp}$ which is added to the $(\alpha_\textsc{tar} + \alpha_{anh})$ part and dominates at frequencies entering the boson peak region.  From an adjustment of the whole data set, the $(\alpha_\textsc{tar} + \alpha_{anh})$-part being fixed at its expected value quoted above, we find $\alpha_{bp} (\omega) = A (\frac{\omega}{2\pi})^4$, with the parameter $A=0.33\pm0.02$~\si{nm^{-1}THz^{-4}} being $T$-independent. The black dashed line in Fig~\ref{fig:results}(a) shows $\alpha_{bp} (\omega)$ while the red and blue solid lines represent the overall total attenuation $\alpha_{tot} (\omega,T)=\alpha_\textsc{tar} + \alpha_{anh} + \alpha_{bp}$ for \SI{300}{K} and \SI{15}{K}, respectively.

Direct demonstration of this highly damped sound wave regime via phonon spectroscopy techniques has been pursued for decades~\cite{Dietsche1979,Zhu1991,Foret1996,Ruffle2003,Ruffle2006,Monaco2009a}. As suggested long ago, the rapid decay of sub-terahertz phonon mean free path must be the cause of the existence of the  universal ``plateau" observed in $\kappa (T)$ of almost all glasses at a few kelvin~\cite{Zeller1971,Pohl2002}. The physical origin of such a strong phonon attenuation regime, which should be closely linked to the glassy state, has long been debated~\cite{Schirmacher2006,Monaco2009b,Buchenau1992,Schober2011,DeGiuli2014,Gelin2016}. The issue is coming back into focus with the emergence of extensive work on the vibrational properties of glasses from computer simulations. Recent investigations on ensembles of computer-samples in different classes of amorphous materials have conclusively demonstrated the existence of soft quasi-localised vibrational modes exhibiting a universal quartic density of states, $\mathcal{D}_\textsc{qlv}(\omega) \sim \omega^4$~\cite{Lerner2016,Mizuno2017,Shimada2018,Lerner2021,Richard2020}. This finding corroborates the long-held prediction of the phenomenological soft potential model (SPM). SPM is an extension of the tunneling model to include soft localised vibrations in terms of soft anharmonic potentials with locally varying parameters~\cite{Buchenau1991,Gil1993,Gurevich2003,Parshin2007}. The QLVs can be understood as soft localised vibrations hybridised with acoustic waves.
They explain the excess modes over the Debye model expectation that form the boson peak~\cite{Lerner2021}. The resonant scattering of sound waves on these soft vibrations would therefore explain the strong acoustic damping regime. The increase $\propto \omega^4$ in acoustic attenuation follows from the quartic increase in ${\cal D}_\textsc{qlv}(\omega)$. On the basis of SPM, LA attenuation is expected as $\alpha_\textsc{spm} = A_\textsc{spm} \left(\frac{\omega}{2 \pi}\right)^4$ with $A_\textsc{spm}=0.23~\si{nm^{-1}THz^{-4}}$ for v-SiO$_2$~\cite{Ruffle2008}, independent of $T$. The latter value is derived from two experimentally accessible quantities, the tunneling strength for LA modes $C_l$ and the density of states of the soft vibrations which can be extracted from low temperature heat capacity data. Our work highlights this strong damping regime along the entire low-frequency side of the boson peak. The obtained $A$-value is in reasonable agreement with the predicted value, $A_\textsc{spm}$. The other prominent theoretical framework, based on the concept of fluctuating elasticity can also explain the observed phenomenology of sound damping and dispersion in glasses~\cite{Schirmacher2006}. Studies indicate that micro-mechanical frustration may play a central role in the emergence of QLVs~\cite{Lerner2018,Kapteijns2021b}. In the case of silica glass, the local oscillatory motions that form QLVs are the famous librations of SiO$_4$ tetrahedra~\cite{Buchenau1984}.

Finally, the LA-linewidth and -velocity data in v-SiO$_2$ at room $T$ using IXS are compared with our results in Fig.~\ref{fig:results}(b)\&(c). Due to resolution limitations, the IXS measurements are constrained to frequencies above $\sim \SI{1}{THz}$, which precludes access to the $\omega\leq\omega_\textsc{bp}$ region. The two $\Gamma(\omega)$ datasets appear to be rather consistent with one another, with the Ioffe-Regel criterion being met at approximately $\omega_\textsc{bp}$. A decrease in the sound velocity accompanies the increase in the damping, see Fig.~\ref{fig:results}(c). However, the negative phase velocity dispersion extracted from the IXS data appears to be somewhat different from that found here. This can be rationalised by considering the fundamental differences between the two techniques. In IXS experiments, the momentum transfer $Q$ (fixed by the instrument setting) defines the wavevector of the phonon under study, the spectrum of which is treated in terms of the damped harmonic oscillator model. This leads to a constant-wavevector phase velocity $v_\textsc{q} =\Omega(Q)/{Q}$ where $\Omega$ is the resonant frequency. It should be emphasised that as the IR limit is approached from below, the acoustic modes are expected to deviate increasingly from plane-waves so that more and more modes contribute to the scattering intensity at the selected $Q$, leading thus to an apparent $v_\textsc{q}$. In our work, on the other hand, the analysis of the phase of acoustic echoes gives access to a constant-frequency phase velocity $v_\omega$. In effective medium theories based on either fluctuating elastic constants~\cite{Schirmacher2006,Schirmacher2007,DeGiuli2014} or resonance on QLVs~\cite{Gurevich2003,Parshin2007}, the response of the disordered system is characterised by $v_\omega$ and not by $v_\textsc{q}$ around the IR-crossover. As shown in~\cite{DeGiuli2014}, $v_\textsc{q}$ and $v_\omega$ are expected to be somewhat different for frequencies approaching this region, in agreement with our observations. Similarly, the propagation of high frequency sound waves is better described by an attenuation coefficient derived from a constant frequency measurement as reported here, rather than by values extracted from IXS constant-$Q$ analysis, $\alpha = \Gamma / v_\textsc{Q} $. It is therefore of great interest to have direct access to the constant-frequency parameters, for both the phase velocity and the scattering length (or attenuation). Above $\omega_\textsc{bp}$, the spectral width observed in IXS is no longer expected to be a true physical attenuation of a sound wave but rather to reflect the superposition of many vibrational modes~\cite{Foret1996,Buchenau2014,Baldi2016}.\\ 

In conclusion, this work has enabled us to disentangle the various contributions to acoustic losses at frequencies entering the boson peak domain, for the prototypical silica glass. As a result of the superposition of several loss mechanisms, {\it effective} frequency power laws are observed for the attenuation coefficient, the exponent of which varies with the ranges of $\omega$ and $T$. This explains the non physical exponent of $\sim 2.9$ observed at \SI{15}{K} in the $150-400 \text{ GHz}$ range. An athermal $\omega^4$-regime for LA attenuation has been identified along the entire low-frequency side of the boson peak. A near quantitative agreement with the SPM prediction is observed. This work represents a significant advance in the field of phonon spectroscopy in disordered media within the challenging terahertz frequency window, using ultrafast optical techniques. Broadband detection configurations offer the unique advantage of direct access to constant-frequency quantities for acoustic damping and phase velocity. This could be highly relevant when vibrational modes cease to resemble plane waves as is expected in glasses in the boson peak region and beyond. Moreover, the technique has the potential to investigate the properties of TA modes, utilising specific implementations. This may be of considerable interest, as it is known that TA modes are dominant in the Debye density of vibrational states.

The authors would like to acknowledge the major contribution of Tsung-Chi Hung on the sample preparation and experimental measurement on the InGaN based samples. This work has been supported by the National Science and Technology Council of Taiwan (Grant No. MOST 110-2112-M-002-033-MY3 and NSTC 113-2112-M-002-019-MY3) and the Agence Nationale de la Recherche (Grant No. ANR-11-BS04-008-01, GlassPhon).


\begin{thebibliography}{76}%
\makeatletter
\providecommand \@ifxundefined [1]{%
 \@ifx{#1\undefined}
}%
\providecommand \@ifnum [1]{%
 \ifnum #1\expandafter \@firstoftwo
 \else \expandafter \@secondoftwo
 \fi
}%
\providecommand \@ifx [1]{%
 \ifx #1\expandafter \@firstoftwo
 \else \expandafter \@secondoftwo
 \fi
}%
\providecommand \natexlab [1]{#1}%
\providecommand \enquote  [1]{``#1''}%
\providecommand \bibnamefont  [1]{#1}%
\providecommand \bibfnamefont [1]{#1}%
\providecommand \citenamefont [1]{#1}%
\providecommand \href@noop [0]{\@secondoftwo}%
\providecommand \href [0]{\begingroup \@sanitize@url \@href}%
\providecommand \@href[1]{\@@startlink{#1}\@@href}%
\providecommand \@@href[1]{\endgroup#1\@@endlink}%
\providecommand \@sanitize@url [0]{\catcode `\\12\catcode `\$12\catcode `\&12\catcode `\#12\catcode `\^12\catcode `\_12\catcode `\%12\relax}%
\providecommand \@@startlink[1]{}%
\providecommand \@@endlink[0]{}%
\providecommand \url  [0]{\begingroup\@sanitize@url \@url }%
\providecommand \@url [1]{\endgroup\@href {#1}{\urlprefix }}%
\providecommand \urlprefix  [0]{URL }%
\providecommand \Eprint [0]{\href }%
\providecommand \doibase [0]{https://doi.org/}%
\providecommand \selectlanguage [0]{\@gobble}%
\providecommand \bibinfo  [0]{\@secondoftwo}%
\providecommand \bibfield  [0]{\@secondoftwo}%
\providecommand \translation [1]{[#1]}%
\providecommand \BibitemOpen [0]{}%
\providecommand \bibitemStop [0]{}%
\providecommand \bibitemNoStop [0]{.\EOS\space}%
\providecommand \EOS [0]{\spacefactor3000\relax}%
\providecommand \BibitemShut  [1]{\csname bibitem#1\endcsname}%
\let\auto@bib@innerbib\@empty
\bibitem [{\citenamefont {Nishikawa}\ \emph {et~al.}(2022)\citenamefont {Nishikawa}, \citenamefont {Ozawa}, \citenamefont {Ikeda}, \citenamefont {Chaudhuri},\ and\ \citenamefont {Berthier}}]{Nishikawa2022}%
  \BibitemOpen
  \bibfield  {author} {\bibinfo {author} {\bibfnamefont {Y.}~\bibnamefont {Nishikawa}}, \bibinfo {author} {\bibfnamefont {M.}~\bibnamefont {Ozawa}}, \bibinfo {author} {\bibfnamefont {A.}~\bibnamefont {Ikeda}}, \bibinfo {author} {\bibfnamefont {P.}~\bibnamefont {Chaudhuri}},\ and\ \bibinfo {author} {\bibfnamefont {L.}~\bibnamefont {Berthier}},\ }\bibfield  {title} {\bibinfo {title} {Relaxation dynamics in the energy landscape of glass-forming liquids},\ }\href {https://doi.org/10.1103/PhysRevX.12.021001} {\bibfield  {journal} {\bibinfo  {journal} {Phys. Rev. X}\ }\textbf {\bibinfo {volume} {12}},\ \bibinfo {pages} {021001} (\bibinfo {year} {2022})}\BibitemShut {NoStop}%
\bibitem [{\citenamefont {Phillips}(1972)}]{Phillips1972}%
  \BibitemOpen
  \bibfield  {author} {\bibinfo {author} {\bibfnamefont {W.~A.}\ \bibnamefont {Phillips}},\ }\bibfield  {title} {\bibinfo {title} {{Tunneling states in amorphous solids}},\ }\href {https://doi.org/10.1007/BF00660072} {\bibfield  {journal} {\bibinfo  {journal} {Journal of Low Temperature Physics}\ }\textbf {\bibinfo {volume} {7}},\ \bibinfo {pages} {351} (\bibinfo {year} {1972})}\BibitemShut {NoStop}%
\bibitem [{\citenamefont {Anderson}\ \emph {et~al.}(1972)\citenamefont {Anderson}, \citenamefont {{Halperin}},\ and\ \citenamefont {{Varma}}}]{Anderson1972}%
  \BibitemOpen
  \bibfield  {author} {\bibinfo {author} {\bibfnamefont {P.~W.}\ \bibnamefont {Anderson}}, \bibinfo {author} {\bibfnamefont {B.~I.}\ \bibnamefont {{Halperin}}},\ and\ \bibinfo {author} {\bibfnamefont {C.~M.}\ \bibnamefont {{Varma}}},\ }\bibfield  {title} {\bibinfo {title} {{Anomalous low-temperature thermal properties of glasses and spin glasses}},\ }\href {https://doi.org/10.1080/14786437208229210} {\bibfield  {journal} {\bibinfo  {journal} {Philosophical Magazine}\ }\textbf {\bibinfo {volume} {25}},\ \bibinfo {pages} {1} (\bibinfo {year} {1972})}\BibitemShut {NoStop}%
\bibitem [{\citenamefont {Phillips}(1987)}]{Phillips1987}%
  \BibitemOpen
  \bibfield  {author} {\bibinfo {author} {\bibfnamefont {W.~A.}\ \bibnamefont {Phillips}},\ }\bibfield  {title} {\bibinfo {title} {Two-level states in glasses},\ }\href@noop {} {\bibfield  {journal} {\bibinfo  {journal} {Rep. Prog. Phys.}\ }\textbf {\bibinfo {volume} {50}},\ \bibinfo {pages} {1657} (\bibinfo {year} {1987})}\BibitemShut {NoStop}%
\bibitem [{\citenamefont {Gilroy}\ and\ \citenamefont {Phillips}(1981)}]{Gilroy1981}%
  \BibitemOpen
  \bibfield  {author} {\bibinfo {author} {\bibfnamefont {K.~S.}\ \bibnamefont {Gilroy}}\ and\ \bibinfo {author} {\bibfnamefont {W.~A.}\ \bibnamefont {Phillips}},\ }\bibfield  {title} {\bibinfo {title} {Asymmetric double-well potential model for structural relaxation processes in amorphous materials},\ }\href@noop {} {\bibfield  {journal} {\bibinfo  {journal} {Philos. Mag.}\ }\textbf {\bibinfo {volume} {43}},\ \bibinfo {pages} {735} (\bibinfo {year} {1981})}\BibitemShut {NoStop}%
\bibitem [{\citenamefont {Pohl}\ \emph {et~al.}(2002)\citenamefont {Pohl}, \citenamefont {Liu},\ and\ \citenamefont {Thompson}}]{Pohl2002}%
  \BibitemOpen
  \bibfield  {author} {\bibinfo {author} {\bibfnamefont {R.~O.}\ \bibnamefont {Pohl}}, \bibinfo {author} {\bibfnamefont {X.}~\bibnamefont {Liu}},\ and\ \bibinfo {author} {\bibfnamefont {E.}~\bibnamefont {Thompson}},\ }\bibfield  {title} {\bibinfo {title} {Low-temperature thermal conductivity and acoustic attenuation in amorphous solids},\ }\href {https://doi.org/10.1103/RevModPhys.74.991} {\bibfield  {journal} {\bibinfo  {journal} {Rev. Mod. Phys.}\ }\textbf {\bibinfo {volume} {74}},\ \bibinfo {pages} {991} (\bibinfo {year} {2002})}\BibitemShut {NoStop}%
\bibitem [{\citenamefont {Damart}\ and\ \citenamefont {Rodney}(2018)}]{Damart2018}%
  \BibitemOpen
  \bibfield  {author} {\bibinfo {author} {\bibfnamefont {T.}~\bibnamefont {Damart}}\ and\ \bibinfo {author} {\bibfnamefont {D.}~\bibnamefont {Rodney}},\ }\bibfield  {title} {\bibinfo {title} {Atomistic study of two-level systems in amorphous silica},\ }\href {https://doi.org/10.1103/PhysRevB.97.014201} {\bibfield  {journal} {\bibinfo  {journal} {Phys. Rev. B}\ }\textbf {\bibinfo {volume} {97}},\ \bibinfo {pages} {014201} (\bibinfo {year} {2018})}\BibitemShut {NoStop}%
\bibitem [{\citenamefont {Rat}\ \emph {et~al.}(2005)\citenamefont {Rat}, \citenamefont {Foret}, \citenamefont {Massiera}, \citenamefont {Vialla}, \citenamefont {Arai}, \citenamefont {Vacher},\ and\ \citenamefont {Courtens}}]{Rat2005}%
  \BibitemOpen
  \bibfield  {author} {\bibinfo {author} {\bibfnamefont {E.}~\bibnamefont {Rat}}, \bibinfo {author} {\bibfnamefont {M.}~\bibnamefont {Foret}}, \bibinfo {author} {\bibfnamefont {G.}~\bibnamefont {Massiera}}, \bibinfo {author} {\bibfnamefont {R.}~\bibnamefont {Vialla}}, \bibinfo {author} {\bibfnamefont {M.}~\bibnamefont {Arai}}, \bibinfo {author} {\bibfnamefont {R.}~\bibnamefont {Vacher}},\ and\ \bibinfo {author} {\bibfnamefont {E.}~\bibnamefont {Courtens}},\ }\bibfield  {title} {\bibinfo {title} {Anharmonic versus relaxational sound damping in glasses. i. brillouin scattering from densified silica},\ }\href {https://doi.org/10.1103/PhysRevB.72.214204} {\bibfield  {journal} {\bibinfo  {journal} {Phys. Rev. B}\ }\textbf {\bibinfo {volume} {72}},\ \bibinfo {pages} {214204} (\bibinfo {year} {2005})}\BibitemShut {NoStop}%
\bibitem [{\citenamefont {Vacher}\ \emph {et~al.}(2005)\citenamefont {Vacher}, \citenamefont {Courtens},\ and\ \citenamefont {Foret}}]{Vacher2005}%
  \BibitemOpen
  \bibfield  {author} {\bibinfo {author} {\bibfnamefont {R.}~\bibnamefont {Vacher}}, \bibinfo {author} {\bibfnamefont {E.}~\bibnamefont {Courtens}},\ and\ \bibinfo {author} {\bibfnamefont {M.}~\bibnamefont {Foret}},\ }\bibfield  {title} {\bibinfo {title} {Anharmonic versus relaxational sound damping in glasses. ii. vitreous silica},\ }\href {https://doi.org/10.1103/PhysRevB.72.214205} {\bibfield  {journal} {\bibinfo  {journal} {Phys. Rev. B}\ }\textbf {\bibinfo {volume} {72}},\ \bibinfo {pages} {214205} (\bibinfo {year} {2005})}\BibitemShut {NoStop}%
\bibitem [{\citenamefont {Zeller}\ and\ \citenamefont {Pohl}(1971)}]{Zeller1971}%
  \BibitemOpen
  \bibfield  {author} {\bibinfo {author} {\bibfnamefont {R.~C.}\ \bibnamefont {Zeller}}\ and\ \bibinfo {author} {\bibfnamefont {R.~O.}\ \bibnamefont {Pohl}},\ }\bibfield  {title} {\bibinfo {title} {Thermal conductivity and specific heat of noncrystalline solids},\ }\href {https://doi.org/10.1103/PhysRevB.4.2029} {\bibfield  {journal} {\bibinfo  {journal} {Phys. Rev. B}\ }\textbf {\bibinfo {volume} {4}},\ \bibinfo {pages} {2029} (\bibinfo {year} {1971})}\BibitemShut {NoStop}%
\bibitem [{\citenamefont {Schirmacher}\ \emph {et~al.}(1998)\citenamefont {Schirmacher}, \citenamefont {Diezemann},\ and\ \citenamefont {Ganter}}]{Schirmacher1998}%
  \BibitemOpen
  \bibfield  {author} {\bibinfo {author} {\bibfnamefont {W.}~\bibnamefont {Schirmacher}}, \bibinfo {author} {\bibfnamefont {G.}~\bibnamefont {Diezemann}},\ and\ \bibinfo {author} {\bibfnamefont {C.}~\bibnamefont {Ganter}},\ }\bibfield  {title} {\bibinfo {title} {Harmonic vibrational excitations in disordered solids and the ``boson peak''},\ }\href {https://doi.org/10.1103/PhysRevLett.81.136} {\bibfield  {journal} {\bibinfo  {journal} {Phys. Rev. Lett.}\ }\textbf {\bibinfo {volume} {81}},\ \bibinfo {pages} {136} (\bibinfo {year} {1998})}\BibitemShut {NoStop}%
\bibitem [{\citenamefont {{Schirmacher, W.}}(2006)}]{Schirmacher2006}%
  \BibitemOpen
  \bibfield  {author} {\bibinfo {author} {\bibnamefont {{Schirmacher, W.}}},\ }\bibfield  {title} {\bibinfo {title} {Thermal conductivity of glassy materials and the "boson peak"},\ }\href {https://doi.org/10.1209/epl/i2005-10471-9} {\bibfield  {journal} {\bibinfo  {journal} {Europhys. Lett.}\ }\textbf {\bibinfo {volume} {73}},\ \bibinfo {pages} {892} (\bibinfo {year} {2006})}\BibitemShut {NoStop}%
\bibitem [{\citenamefont {Gelin}\ \emph {et~al.}(2016)\citenamefont {Gelin}, \citenamefont {Tanaka},\ and\ \citenamefont {Lema\^{\i}tre}}]{Gelin2016}%
  \BibitemOpen
  \bibfield  {author} {\bibinfo {author} {\bibfnamefont {S.}~\bibnamefont {Gelin}}, \bibinfo {author} {\bibfnamefont {H.}~\bibnamefont {Tanaka}},\ and\ \bibinfo {author} {\bibfnamefont {A.}~\bibnamefont {Lema\^{\i}tre}},\ }\bibfield  {title} {\bibinfo {title} {Anomalous phonon scattering and elastic correlations in amorphous solids},\ }\href {https://doi.org/https://doi.org/10.1038/nmat4736} {\bibfield  {journal} {\bibinfo  {journal} {Nature Materials}\ }\textbf {\bibinfo {volume} {15}},\ \bibinfo {pages} {1177} (\bibinfo {year} {2016})}\BibitemShut {NoStop}%
\bibitem [{\citenamefont {L\'eonforte}\ \emph {et~al.}(2006)\citenamefont {L\'eonforte}, \citenamefont {Tanguy}, \citenamefont {Wittmer},\ and\ \citenamefont {Barrat}}]{Leonforte2006}%
  \BibitemOpen
  \bibfield  {author} {\bibinfo {author} {\bibfnamefont {F.}~\bibnamefont {L\'eonforte}}, \bibinfo {author} {\bibfnamefont {A.}~\bibnamefont {Tanguy}}, \bibinfo {author} {\bibfnamefont {J.~P.}\ \bibnamefont {Wittmer}},\ and\ \bibinfo {author} {\bibfnamefont {J.-L.}\ \bibnamefont {Barrat}},\ }\bibfield  {title} {\bibinfo {title} {Inhomogeneous elastic response of silica glass},\ }\href {https://doi.org/10.1103/PhysRevLett.97.055501} {\bibfield  {journal} {\bibinfo  {journal} {Phys. Rev. Lett.}\ }\textbf {\bibinfo {volume} {97}},\ \bibinfo {pages} {055501} (\bibinfo {year} {2006})}\BibitemShut {NoStop}%
\bibitem [{\citenamefont {Caroli}\ and\ \citenamefont {Lema\^{\i}tre}(2019)}]{Caroli2019}%
  \BibitemOpen
  \bibfield  {author} {\bibinfo {author} {\bibfnamefont {C.}~\bibnamefont {Caroli}}\ and\ \bibinfo {author} {\bibfnamefont {A.}~\bibnamefont {Lema\^{\i}tre}},\ }\bibfield  {title} {\bibinfo {title} {Fluctuating elasticity fails to capture anomalous sound scattering in amorphous solids},\ }\href {https://doi.org/10.1103/PhysRevLett.123.055501} {\bibfield  {journal} {\bibinfo  {journal} {Phys. Rev. Lett.}\ }\textbf {\bibinfo {volume} {123}},\ \bibinfo {pages} {055501} (\bibinfo {year} {2019})}\BibitemShut {NoStop}%
\bibitem [{\citenamefont {Tanguy}\ \emph {et~al.}(2002)\citenamefont {Tanguy}, \citenamefont {Wittmer}, \citenamefont {Leonforte},\ and\ \citenamefont {Barrat}}]{Tanguy2002}%
  \BibitemOpen
  \bibfield  {author} {\bibinfo {author} {\bibfnamefont {A.}~\bibnamefont {Tanguy}}, \bibinfo {author} {\bibfnamefont {J.~P.}\ \bibnamefont {Wittmer}}, \bibinfo {author} {\bibfnamefont {F.}~\bibnamefont {Leonforte}},\ and\ \bibinfo {author} {\bibfnamefont {J.-L.}\ \bibnamefont {Barrat}},\ }\bibfield  {title} {\bibinfo {title} {Continuum limit of amorphous elastic bodies: A finite-size study of low-frequency harmonic vibrations},\ }\href {https://doi.org/10.1103/PhysRevB.66.174205} {\bibfield  {journal} {\bibinfo  {journal} {Phys. Rev. B}\ }\textbf {\bibinfo {volume} {66}},\ \bibinfo {pages} {174205} (\bibinfo {year} {2002})}\BibitemShut {NoStop}%
\bibitem [{\citenamefont {Szamel}\ and\ \citenamefont {Flenner}(2022)}]{Szamel2022}%
  \BibitemOpen
  \bibfield  {author} {\bibinfo {author} {\bibfnamefont {G.}~\bibnamefont {Szamel}}\ and\ \bibinfo {author} {\bibfnamefont {E.}~\bibnamefont {Flenner}},\ }\bibfield  {title} {\bibinfo {title} {{Microscopic analysis of sound attenuation in low-temperature amorphous solids reveals quantitative importance of non-affine effects}},\ }\href {https://doi.org/10.1063/5.0085199} {\bibfield  {journal} {\bibinfo  {journal} {The Journal of Chemical Physics}\ }\textbf {\bibinfo {volume} {156}},\ \bibinfo {pages} {144502} (\bibinfo {year} {2022})},\ \Eprint {https://arxiv.org/abs/https://pubs.aip.org/aip/jcp/article-pdf/doi/10.1063/5.0085199/16539919/144502\_1\_online.pdf} {https://pubs.aip.org/aip/jcp/article-pdf/doi/10.1063/5.0085199/16539919/144502\_1\_online.pdf} \BibitemShut {NoStop}%
\bibitem [{\citenamefont {Baggioli}\ and\ \citenamefont {Zaccone}(2022)}]{Baggioli2022}%
  \BibitemOpen
  \bibfield  {author} {\bibinfo {author} {\bibfnamefont {M.}~\bibnamefont {Baggioli}}\ and\ \bibinfo {author} {\bibfnamefont {A.}~\bibnamefont {Zaccone}},\ }\bibfield  {title} {\bibinfo {title} {Theory of sound attenuation in amorphous solids from nonaffine motions},\ }\href {https://doi.org/10.1088/1361-648X/ac5d8b} {\bibfield  {journal} {\bibinfo  {journal} {Journal of Physics: Condensed Matter}\ }\textbf {\bibinfo {volume} {34}},\ \bibinfo {pages} {215401} (\bibinfo {year} {2022})}\BibitemShut {NoStop}%
\bibitem [{\citenamefont {Gurevich}\ \emph {et~al.}(2003)\citenamefont {Gurevich}, \citenamefont {Parshin},\ and\ \citenamefont {Schober}}]{Gurevich2003}%
  \BibitemOpen
  \bibfield  {author} {\bibinfo {author} {\bibfnamefont {V.~L.}\ \bibnamefont {Gurevich}}, \bibinfo {author} {\bibfnamefont {D.~A.}\ \bibnamefont {Parshin}},\ and\ \bibinfo {author} {\bibfnamefont {H.~R.}\ \bibnamefont {Schober}},\ }\bibfield  {title} {\bibinfo {title} {Anharmonicity, vibrational instability, and the boson peak in glasses},\ }\href {https://doi.org/10.1103/PhysRevB.67.094203} {\bibfield  {journal} {\bibinfo  {journal} {Phys. Rev. B}\ }\textbf {\bibinfo {volume} {67}},\ \bibinfo {pages} {094203} (\bibinfo {year} {2003})}\BibitemShut {NoStop}%
\bibitem [{\citenamefont {Parshin}\ \emph {et~al.}(2007)\citenamefont {Parshin}, \citenamefont {Schober},\ and\ \citenamefont {Gurevich}}]{Parshin2007}%
  \BibitemOpen
  \bibfield  {author} {\bibinfo {author} {\bibfnamefont {D.~A.}\ \bibnamefont {Parshin}}, \bibinfo {author} {\bibfnamefont {H.~R.}\ \bibnamefont {Schober}},\ and\ \bibinfo {author} {\bibfnamefont {V.~L.}\ \bibnamefont {Gurevich}},\ }\bibfield  {title} {\bibinfo {title} {Vibrational instability, two-level systems, and the boson peak in glasses},\ }\href {https://doi.org/10.1103/PhysRevB.76.064206} {\bibfield  {journal} {\bibinfo  {journal} {Phys. Rev. B}\ }\textbf {\bibinfo {volume} {76}},\ \bibinfo {pages} {064206} (\bibinfo {year} {2007})}\BibitemShut {NoStop}%
\bibitem [{\citenamefont {Lerner}\ \emph {et~al.}(2016)\citenamefont {Lerner}, \citenamefont {D\"uring},\ and\ \citenamefont {Bouchbinder}}]{Lerner2016}%
  \BibitemOpen
  \bibfield  {author} {\bibinfo {author} {\bibfnamefont {E.}~\bibnamefont {Lerner}}, \bibinfo {author} {\bibfnamefont {G.}~\bibnamefont {D\"uring}},\ and\ \bibinfo {author} {\bibfnamefont {E.}~\bibnamefont {Bouchbinder}},\ }\bibfield  {title} {\bibinfo {title} {Statistics and properties of low-frequency vibrational modes in structural glasses},\ }\href {https://doi.org/10.1103/PhysRevLett.117.035501} {\bibfield  {journal} {\bibinfo  {journal} {Phys. Rev. Lett.}\ }\textbf {\bibinfo {volume} {117}},\ \bibinfo {pages} {035501} (\bibinfo {year} {2016})}\BibitemShut {NoStop}%
\bibitem [{\citenamefont {Shimada}\ \emph {et~al.}(2018)\citenamefont {Shimada}, \citenamefont {Mizuno}, \citenamefont {Wyart},\ and\ \citenamefont {Ikeda}}]{Shimada2018}%
  \BibitemOpen
  \bibfield  {author} {\bibinfo {author} {\bibfnamefont {M.}~\bibnamefont {Shimada}}, \bibinfo {author} {\bibfnamefont {H.}~\bibnamefont {Mizuno}}, \bibinfo {author} {\bibfnamefont {M.}~\bibnamefont {Wyart}},\ and\ \bibinfo {author} {\bibfnamefont {A.}~\bibnamefont {Ikeda}},\ }\bibfield  {title} {\bibinfo {title} {Spatial structure of quasilocalized vibrations in nearly jammed amorphous solids},\ }\href {https://doi.org/10.1103/PhysRevE.98.060901} {\bibfield  {journal} {\bibinfo  {journal} {Phys. Rev. E}\ }\textbf {\bibinfo {volume} {98}},\ \bibinfo {pages} {060901} (\bibinfo {year} {2018})}\BibitemShut {NoStop}%
\bibitem [{\citenamefont {Wang}\ \emph {et~al.}(2019)\citenamefont {Wang}, \citenamefont {Berthier}, \citenamefont {Flenner}, \citenamefont {Guan},\ and\ \citenamefont {Szamel}}]{Wang2019}%
  \BibitemOpen
  \bibfield  {author} {\bibinfo {author} {\bibfnamefont {L.}~\bibnamefont {Wang}}, \bibinfo {author} {\bibfnamefont {L.}~\bibnamefont {Berthier}}, \bibinfo {author} {\bibfnamefont {E.}~\bibnamefont {Flenner}}, \bibinfo {author} {\bibfnamefont {P.}~\bibnamefont {Guan}},\ and\ \bibinfo {author} {\bibfnamefont {G.}~\bibnamefont {Szamel}},\ }\bibfield  {title} {\bibinfo {title} {Sound attenuation in stable glasses},\ }\href {https://doi.org/10.1039/C9SM01092K} {\bibfield  {journal} {\bibinfo  {journal} {Soft Matter}\ }\textbf {\bibinfo {volume} {15}},\ \bibinfo {pages} {7018} (\bibinfo {year} {2019})}\BibitemShut {NoStop}%
\bibitem [{\citenamefont {DeGiuli}\ \emph {et~al.}(2014)\citenamefont {DeGiuli}, \citenamefont {Laversanne-Finot}, \citenamefont {During}, \citenamefont {Lerner},\ and\ \citenamefont {Wyart}}]{DeGiuli2014}%
  \BibitemOpen
  \bibfield  {author} {\bibinfo {author} {\bibfnamefont {E.}~\bibnamefont {DeGiuli}}, \bibinfo {author} {\bibfnamefont {A.}~\bibnamefont {Laversanne-Finot}}, \bibinfo {author} {\bibfnamefont {G.}~\bibnamefont {During}}, \bibinfo {author} {\bibfnamefont {E.}~\bibnamefont {Lerner}},\ and\ \bibinfo {author} {\bibfnamefont {M.}~\bibnamefont {Wyart}},\ }\bibfield  {title} {\bibinfo {title} {Effects of coordination and pressure on sound attenuation, boson peak and elasticity in amorphous solids},\ }\href {https://doi.org/10.1039/C4SM00561A} {\bibfield  {journal} {\bibinfo  {journal} {Soft Matter}\ }\textbf {\bibinfo {volume} {10}},\ \bibinfo {pages} {5628} (\bibinfo {year} {2014})}\BibitemShut {NoStop}%
\bibitem [{\citenamefont {Conyuh}\ and\ \citenamefont {Beltukov}(2021)}]{Conyuh2021}%
  \BibitemOpen
  \bibfield  {author} {\bibinfo {author} {\bibfnamefont {D.~A.}\ \bibnamefont {Conyuh}}\ and\ \bibinfo {author} {\bibfnamefont {Y.~M.}\ \bibnamefont {Beltukov}},\ }\bibfield  {title} {\bibinfo {title} {Random matrix approach to the boson peak and ioffe-regel criterion in amorphous solids},\ }\href {https://doi.org/10.1103/PhysRevB.103.104204} {\bibfield  {journal} {\bibinfo  {journal} {Phys. Rev. B}\ }\textbf {\bibinfo {volume} {103}},\ \bibinfo {pages} {104204} (\bibinfo {year} {2021})}\BibitemShut {NoStop}%
\bibitem [{\citenamefont {Moriel}\ \emph {et~al.}(2019)\citenamefont {Moriel}, \citenamefont {Kapteijns}, \citenamefont {Rainone}, \citenamefont {Zylberg}, \citenamefont {Lerner},\ and\ \citenamefont {Bouchbinder}}]{Moriel2019}%
  \BibitemOpen
  \bibfield  {author} {\bibinfo {author} {\bibfnamefont {A.}~\bibnamefont {Moriel}}, \bibinfo {author} {\bibfnamefont {G.}~\bibnamefont {Kapteijns}}, \bibinfo {author} {\bibfnamefont {C.}~\bibnamefont {Rainone}}, \bibinfo {author} {\bibfnamefont {J.}~\bibnamefont {Zylberg}}, \bibinfo {author} {\bibfnamefont {E.}~\bibnamefont {Lerner}},\ and\ \bibinfo {author} {\bibfnamefont {E.}~\bibnamefont {Bouchbinder}},\ }\bibfield  {title} {\bibinfo {title} {{Wave attenuation in glasses: Rayleigh and generalized-Rayleigh scattering scaling}},\ }\href {https://doi.org/10.1063/1.5111192} {\bibfield  {journal} {\bibinfo  {journal} {The Journal of Chemical Physics}\ }\textbf {\bibinfo {volume} {151}},\ \bibinfo {pages} {104503} (\bibinfo {year} {2019})},\ \Eprint {https://arxiv.org/abs/https://pubs.aip.org/aip/jcp/article-pdf/doi/10.1063/1.5111192/13604680/104503\_1\_online.pdf} {https://pubs.aip.org/aip/jcp/article-pdf/doi/10.1063/1.5111192/13604680/104503\_1\_online.pdf} \BibitemShut {NoStop}%
\bibitem [{\citenamefont {Ruffl\'e}\ \emph {et~al.}(2003)\citenamefont {Ruffl\'e}, \citenamefont {Foret}, \citenamefont {Courtens}, \citenamefont {Vacher},\ and\ \citenamefont {Monaco}}]{Ruffle2003}%
  \BibitemOpen
  \bibfield  {author} {\bibinfo {author} {\bibfnamefont {B.}~\bibnamefont {Ruffl\'e}}, \bibinfo {author} {\bibfnamefont {M.}~\bibnamefont {Foret}}, \bibinfo {author} {\bibfnamefont {E.}~\bibnamefont {Courtens}}, \bibinfo {author} {\bibfnamefont {R.}~\bibnamefont {Vacher}},\ and\ \bibinfo {author} {\bibfnamefont {G.}~\bibnamefont {Monaco}},\ }\bibfield  {title} {\bibinfo {title} {Observation of the onset of strong scattering on high frequency acoustic phonons in densified silica glass},\ }\href {https://doi.org/10.1103/PhysRevLett.90.095502} {\bibfield  {journal} {\bibinfo  {journal} {Phys. Rev. Lett.}\ }\textbf {\bibinfo {volume} {90}},\ \bibinfo {pages} {095502} (\bibinfo {year} {2003})}\BibitemShut {NoStop}%
\bibitem [{\citenamefont {Ruffl\'e}\ \emph {et~al.}(2006)\citenamefont {Ruffl\'e}, \citenamefont {Guimbreti\`ere}, \citenamefont {Courtens}, \citenamefont {Vacher},\ and\ \citenamefont {Monaco}}]{Ruffle2006}%
  \BibitemOpen
  \bibfield  {author} {\bibinfo {author} {\bibfnamefont {B.}~\bibnamefont {Ruffl\'e}}, \bibinfo {author} {\bibfnamefont {G.}~\bibnamefont {Guimbreti\`ere}}, \bibinfo {author} {\bibfnamefont {E.}~\bibnamefont {Courtens}}, \bibinfo {author} {\bibfnamefont {R.}~\bibnamefont {Vacher}},\ and\ \bibinfo {author} {\bibfnamefont {G.}~\bibnamefont {Monaco}},\ }\bibfield  {title} {\bibinfo {title} {Glass-specific behavior in the damping of acousticlike vibrations},\ }\href {https://doi.org/10.1103/PhysRevLett.96.045502} {\bibfield  {journal} {\bibinfo  {journal} {Phys. Rev. Lett.}\ }\textbf {\bibinfo {volume} {96}},\ \bibinfo {pages} {045502} (\bibinfo {year} {2006})}\BibitemShut {NoStop}%
\bibitem [{\citenamefont {Monaco}\ \emph {et~al.}(2009)\citenamefont {Monaco}, \citenamefont {Giordano},\ and\ \citenamefont {Stanley}}]{Monaco2009a}%
  \BibitemOpen
  \bibfield  {author} {\bibinfo {author} {\bibfnamefont {G.}~\bibnamefont {Monaco}}, \bibinfo {author} {\bibfnamefont {V.~M.}\ \bibnamefont {Giordano}},\ and\ \bibinfo {author} {\bibfnamefont {H.~E.}\ \bibnamefont {Stanley}},\ }\bibfield  {title} {\bibinfo {title} {Breakdown of the debye approximation for the acoustic modes with nanometric wavelengths in glasses},\ }\href {http://www.jstor.org/stable/40428431} {\bibfield  {journal} {\bibinfo  {journal} {Proceedings of the National Academy of Sciences of the United States of America}\ }\textbf {\bibinfo {volume} {106}},\ \bibinfo {pages} {3659} (\bibinfo {year} {2009})}\BibitemShut {NoStop}%
\bibitem [{\citenamefont {Baldi}\ \emph {et~al.}(2010)\citenamefont {Baldi}, \citenamefont {Giordano}, \citenamefont {Monaco},\ and\ \citenamefont {Ruta}}]{Baldi2010}%
  \BibitemOpen
  \bibfield  {author} {\bibinfo {author} {\bibfnamefont {G.}~\bibnamefont {Baldi}}, \bibinfo {author} {\bibfnamefont {V.~M.}\ \bibnamefont {Giordano}}, \bibinfo {author} {\bibfnamefont {G.}~\bibnamefont {Monaco}},\ and\ \bibinfo {author} {\bibfnamefont {B.}~\bibnamefont {Ruta}},\ }\bibfield  {title} {\bibinfo {title} {Sound attenuation at terahertz frequencies and the boson peak of vitreous silica},\ }\href {https://doi.org/10.1103/PhysRevLett.104.195501} {\bibfield  {journal} {\bibinfo  {journal} {Phys. Rev. Lett.}\ }\textbf {\bibinfo {volume} {104}},\ \bibinfo {pages} {195501} (\bibinfo {year} {2010})}\BibitemShut {NoStop}%
\bibitem [{\citenamefont {Huynh}\ \emph {et~al.}(2006)\citenamefont {Huynh}, \citenamefont {Lanzillotti-Kimura}, \citenamefont {Jusserand}, \citenamefont {Perrin}, \citenamefont {Fainstein}, \citenamefont {Pascual-Winter}, \citenamefont {Peronne},\ and\ \citenamefont {Lema\^{\i}tre}}]{Huynh2006}%
  \BibitemOpen
  \bibfield  {author} {\bibinfo {author} {\bibfnamefont {A.}~\bibnamefont {Huynh}}, \bibinfo {author} {\bibfnamefont {N.~D.}\ \bibnamefont {Lanzillotti-Kimura}}, \bibinfo {author} {\bibfnamefont {B.}~\bibnamefont {Jusserand}}, \bibinfo {author} {\bibfnamefont {B.}~\bibnamefont {Perrin}}, \bibinfo {author} {\bibfnamefont {A.}~\bibnamefont {Fainstein}}, \bibinfo {author} {\bibfnamefont {M.~F.}\ \bibnamefont {Pascual-Winter}}, \bibinfo {author} {\bibfnamefont {E.}~\bibnamefont {Peronne}},\ and\ \bibinfo {author} {\bibfnamefont {A.}~\bibnamefont {Lema\^{\i}tre}},\ }\bibfield  {title} {\bibinfo {title} {Subterahertz phonon dynamics in acoustic nanocavities},\ }\href {https://doi.org/10.1103/PhysRevLett.97.115502} {\bibfield  {journal} {\bibinfo  {journal} {Phys. Rev. Lett.}\ }\textbf {\bibinfo {volume} {97}},\ \bibinfo {pages} {115502} (\bibinfo {year} {2006})}\BibitemShut {NoStop}%
\bibitem [{\citenamefont {Maznev}\ \emph {et~al.}(2012)\citenamefont {Maznev}, \citenamefont {Manke}, \citenamefont {Lin}, \citenamefont {Nelson}, \citenamefont {Sun},\ and\ \citenamefont {Chyi}}]{Maznev2012}%
  \BibitemOpen
  \bibfield  {author} {\bibinfo {author} {\bibfnamefont {A.}~\bibnamefont {Maznev}}, \bibinfo {author} {\bibfnamefont {K.~J.}\ \bibnamefont {Manke}}, \bibinfo {author} {\bibfnamefont {K.-H.}\ \bibnamefont {Lin}}, \bibinfo {author} {\bibfnamefont {K.~A.}\ \bibnamefont {Nelson}}, \bibinfo {author} {\bibfnamefont {C.-K.}\ \bibnamefont {Sun}},\ and\ \bibinfo {author} {\bibfnamefont {J.-I.}\ \bibnamefont {Chyi}},\ }\bibfield  {title} {\bibinfo {title} {Broadband terahertz ultrasonic transducer based on a laser-driven piezoelectric semiconductor superlattice},\ }\href {https://doi.org/https://doi.org/10.1016/j.ultras.2011.07.007} {\bibfield  {journal} {\bibinfo  {journal} {Ultrasonics}\ }\textbf {\bibinfo {volume} {52}},\ \bibinfo {pages} {1 } (\bibinfo {year} {2012})}\BibitemShut {NoStop}%
\bibitem [{\citenamefont {Devos}\ \emph {et~al.}(2008)\citenamefont {Devos}, \citenamefont {Foret}, \citenamefont {Ayrinhac}, \citenamefont {Emery},\ and\ \citenamefont {Ruffl\'e}}]{Devos2008}%
  \BibitemOpen
  \bibfield  {author} {\bibinfo {author} {\bibfnamefont {A.}~\bibnamefont {Devos}}, \bibinfo {author} {\bibfnamefont {M.}~\bibnamefont {Foret}}, \bibinfo {author} {\bibfnamefont {S.}~\bibnamefont {Ayrinhac}}, \bibinfo {author} {\bibfnamefont {P.}~\bibnamefont {Emery}},\ and\ \bibinfo {author} {\bibfnamefont {B.}~\bibnamefont {Ruffl\'e}},\ }\bibfield  {title} {\bibinfo {title} {Hypersound damping in vitreous silica measured by picosecond acoustics},\ }\href {https://doi.org/10.1103/PhysRevB.77.100201} {\bibfield  {journal} {\bibinfo  {journal} {Phys. Rev. B}\ }\textbf {\bibinfo {volume} {77}},\ \bibinfo {pages} {100201} (\bibinfo {year} {2008})}\BibitemShut {NoStop}%
\bibitem [{\citenamefont {Klieber}\ \emph {et~al.}(2011)\citenamefont {Klieber}, \citenamefont {Peronne}, \citenamefont {Katayama}, \citenamefont {Choi}, \citenamefont {Yamaguchi}, \citenamefont {Pezeril},\ and\ \citenamefont {Nelson}}]{Klieber2011}%
  \BibitemOpen
  \bibfield  {author} {\bibinfo {author} {\bibfnamefont {C.}~\bibnamefont {Klieber}}, \bibinfo {author} {\bibfnamefont {E.}~\bibnamefont {Peronne}}, \bibinfo {author} {\bibfnamefont {K.}~\bibnamefont {Katayama}}, \bibinfo {author} {\bibfnamefont {J.}~\bibnamefont {Choi}}, \bibinfo {author} {\bibfnamefont {M.}~\bibnamefont {Yamaguchi}}, \bibinfo {author} {\bibfnamefont {T.}~\bibnamefont {Pezeril}},\ and\ \bibinfo {author} {\bibfnamefont {K.~a.}\ \bibnamefont {Nelson}},\ }\bibfield  {title} {\bibinfo {title} {{Narrow-band acoustic attenuation measurements in vitreous silica at frequencies between 20 and 400 GHz}},\ }\href {https://doi.org/10.1063/1.3595275} {\bibfield  {journal} {\bibinfo  {journal} {Applied Physics Letters}\ }\textbf {\bibinfo {volume} {98}},\ \bibinfo {pages} {211908} (\bibinfo {year} {2011})}\BibitemShut {NoStop}%
\bibitem [{\citenamefont {Foglia}\ \emph {et~al.}(2023)\citenamefont {Foglia}, \citenamefont {Mincigrucci}, \citenamefont {Maznev}, \citenamefont {Baldi}, \citenamefont {Capotondi}, \citenamefont {Caporaletti}, \citenamefont {Comin}, \citenamefont {{De Angelis}}, \citenamefont {Duncan}, \citenamefont {Fainozzi}, \citenamefont {Kurdi}, \citenamefont {Li}, \citenamefont {Martinelli}, \citenamefont {Masciovecchio}, \citenamefont {Monaco}, \citenamefont {Milloch}, \citenamefont {Nelson}, \citenamefont {Occhialini}, \citenamefont {Pancaldi}, \citenamefont {Pedersoli}, \citenamefont {Pelli-Cresi}, \citenamefont {Simoncig}, \citenamefont {Travasso}, \citenamefont {Wehinger}, \citenamefont {Zanatta},\ and\ \citenamefont {Bencivenga}}]{Foglia2023}%
  \BibitemOpen
  \bibfield  {author} {\bibinfo {author} {\bibfnamefont {L.}~\bibnamefont {Foglia}}, \bibinfo {author} {\bibfnamefont {R.}~\bibnamefont {Mincigrucci}}, \bibinfo {author} {\bibfnamefont {A.~A.}\ \bibnamefont {Maznev}}, \bibinfo {author} {\bibfnamefont {G.}~\bibnamefont {Baldi}}, \bibinfo {author} {\bibfnamefont {F.}~\bibnamefont {Capotondi}}, \bibinfo {author} {\bibfnamefont {F.}~\bibnamefont {Caporaletti}}, \bibinfo {author} {\bibfnamefont {R.}~\bibnamefont {Comin}}, \bibinfo {author} {\bibfnamefont {D.}~\bibnamefont {{De Angelis}}}, \bibinfo {author} {\bibfnamefont {R.~A.}\ \bibnamefont {Duncan}}, \bibinfo {author} {\bibfnamefont {D.}~\bibnamefont {Fainozzi}}, \bibinfo {author} {\bibfnamefont {G.}~\bibnamefont {Kurdi}}, \bibinfo {author} {\bibfnamefont {J.}~\bibnamefont {Li}}, \bibinfo {author} {\bibfnamefont {A.}~\bibnamefont {Martinelli}}, \bibinfo {author} {\bibfnamefont {C.}~\bibnamefont {Masciovecchio}}, \bibinfo {author} {\bibfnamefont {G.}~\bibnamefont {Monaco}}, \bibinfo {author} {\bibfnamefont
  {A.}~\bibnamefont {Milloch}}, \bibinfo {author} {\bibfnamefont {K.~A.}\ \bibnamefont {Nelson}}, \bibinfo {author} {\bibfnamefont {C.~A.}\ \bibnamefont {Occhialini}}, \bibinfo {author} {\bibfnamefont {M.}~\bibnamefont {Pancaldi}}, \bibinfo {author} {\bibfnamefont {E.}~\bibnamefont {Pedersoli}}, \bibinfo {author} {\bibfnamefont {J.~S.}\ \bibnamefont {Pelli-Cresi}}, \bibinfo {author} {\bibfnamefont {A.}~\bibnamefont {Simoncig}}, \bibinfo {author} {\bibfnamefont {F.}~\bibnamefont {Travasso}}, \bibinfo {author} {\bibfnamefont {B.}~\bibnamefont {Wehinger}}, \bibinfo {author} {\bibfnamefont {M.}~\bibnamefont {Zanatta}},\ and\ \bibinfo {author} {\bibfnamefont {F.}~\bibnamefont {Bencivenga}},\ }\bibfield  {title} {\bibinfo {title} {Extreme ultraviolet transient gratings: A tool for nanoscale photoacoustics.},\ }\href {https://doi.org/10.1016/j.pacs.2023.100453} {\bibfield  {journal} {\bibinfo  {journal} {Photoacoustics}\ }\textbf {\bibinfo {volume} {29}},\ \bibinfo {pages} {100453} (\bibinfo {year}
  {2023})}\BibitemShut {NoStop}%
\bibitem [{\citenamefont {Fainozzi}\ \emph {et~al.}(2024)\citenamefont {Fainozzi}, \citenamefont {Foglia}, \citenamefont {Khatu}, \citenamefont {Masciovecchio}, \citenamefont {Mincigrucci}, \citenamefont {Paltanin},\ and\ \citenamefont {Bencivenga}}]{Fainozzi2024}%
  \BibitemOpen
  \bibfield  {author} {\bibinfo {author} {\bibfnamefont {D.}~\bibnamefont {Fainozzi}}, \bibinfo {author} {\bibfnamefont {L.}~\bibnamefont {Foglia}}, \bibinfo {author} {\bibfnamefont {N.~N.}\ \bibnamefont {Khatu}}, \bibinfo {author} {\bibfnamefont {C.}~\bibnamefont {Masciovecchio}}, \bibinfo {author} {\bibfnamefont {R.}~\bibnamefont {Mincigrucci}}, \bibinfo {author} {\bibfnamefont {E.}~\bibnamefont {Paltanin}},\ and\ \bibinfo {author} {\bibfnamefont {F.}~\bibnamefont {Bencivenga}},\ }\bibfield  {title} {\bibinfo {title} {Stimulated brillouin scattering in the time domain at $1\text{ }\text{ }{\mathrm{nm}}^{\ensuremath{-}1}$ wave vector},\ }\href {https://doi.org/10.1103/PhysRevLett.132.033802} {\bibfield  {journal} {\bibinfo  {journal} {Phys. Rev. Lett.}\ }\textbf {\bibinfo {volume} {132}},\ \bibinfo {pages} {033802} (\bibinfo {year} {2024})}\BibitemShut {NoStop}%
\bibitem [{\citenamefont {Huynh}\ \emph {et~al.}(2008)\citenamefont {Huynh}, \citenamefont {Perrin}, \citenamefont {Lanzillotti-Kimura}, \citenamefont {Jusserand}, \citenamefont {Fainstein},\ and\ \citenamefont {Lema\^{\i}tre}}]{Huynh2008}%
  \BibitemOpen
  \bibfield  {author} {\bibinfo {author} {\bibfnamefont {A.}~\bibnamefont {Huynh}}, \bibinfo {author} {\bibfnamefont {B.}~\bibnamefont {Perrin}}, \bibinfo {author} {\bibfnamefont {N.~D.}\ \bibnamefont {Lanzillotti-Kimura}}, \bibinfo {author} {\bibfnamefont {B.}~\bibnamefont {Jusserand}}, \bibinfo {author} {\bibfnamefont {A.}~\bibnamefont {Fainstein}},\ and\ \bibinfo {author} {\bibfnamefont {A.}~\bibnamefont {Lema\^{\i}tre}},\ }\bibfield  {title} {\bibinfo {title} {Subterahertz monochromatic acoustic wave propagation using semiconductor superlattices as transducers},\ }\href {https://doi.org/10.1103/PhysRevB.78.233302} {\bibfield  {journal} {\bibinfo  {journal} {Phys. Rev. B}\ }\textbf {\bibinfo {volume} {78}},\ \bibinfo {pages} {233302} (\bibinfo {year} {2008})}\BibitemShut {NoStop}%
\bibitem [{\citenamefont {Huynh}\ \emph {et~al.}(2015)\citenamefont {Huynh}, \citenamefont {Perrin},\ and\ \citenamefont {Lema\^{\i}tre}}]{Huynh2015}%
  \BibitemOpen
  \bibfield  {author} {\bibinfo {author} {\bibfnamefont {A.}~\bibnamefont {Huynh}}, \bibinfo {author} {\bibfnamefont {B.}~\bibnamefont {Perrin}},\ and\ \bibinfo {author} {\bibfnamefont {A.}~\bibnamefont {Lema\^{\i}tre}},\ }\bibfield  {title} {\bibinfo {title} {Semiconductor superlattices: A tool for terahertz acoustics},\ }\href {https://doi.org/https://doi.org/10.1016/j.ultras.2014.07.009} {\bibfield  {journal} {\bibinfo  {journal} {Ultrasonics}\ }\textbf {\bibinfo {volume} {56}},\ \bibinfo {pages} {66 } (\bibinfo {year} {2015})}\BibitemShut {NoStop}%
\bibitem [{\citenamefont {Huynh}\ \emph {et~al.}(2017)\citenamefont {Huynh}, \citenamefont {P\'eronne}, \citenamefont {Gingreau}, \citenamefont {Lafosse}, \citenamefont {Lema\^{\i}tre}, \citenamefont {Perrin}, \citenamefont {Vacher}, \citenamefont {Ruffl\'e},\ and\ \citenamefont {Foret}}]{Huynh2017}%
  \BibitemOpen
  \bibfield  {author} {\bibinfo {author} {\bibfnamefont {A.}~\bibnamefont {Huynh}}, \bibinfo {author} {\bibfnamefont {E.}~\bibnamefont {P\'eronne}}, \bibinfo {author} {\bibfnamefont {C.}~\bibnamefont {Gingreau}}, \bibinfo {author} {\bibfnamefont {X.}~\bibnamefont {Lafosse}}, \bibinfo {author} {\bibfnamefont {A.}~\bibnamefont {Lema\^{\i}tre}}, \bibinfo {author} {\bibfnamefont {B.}~\bibnamefont {Perrin}}, \bibinfo {author} {\bibfnamefont {R.}~\bibnamefont {Vacher}}, \bibinfo {author} {\bibfnamefont {B.}~\bibnamefont {Ruffl\'e}},\ and\ \bibinfo {author} {\bibfnamefont {M.}~\bibnamefont {Foret}},\ }\bibfield  {title} {\bibinfo {title} {Temperature dependence of hypersound attenuation in silica films via picosecond acoustics},\ }\href {https://doi.org/10.1103/PhysRevB.96.174206} {\bibfield  {journal} {\bibinfo  {journal} {Phys. Rev. B}\ }\textbf {\bibinfo {volume} {96}},\ \bibinfo {pages} {174206} (\bibinfo {year} {2017})}\BibitemShut {NoStop}%
\bibitem [{\citenamefont {Dietsche}\ and\ \citenamefont {Kinder}(1979)}]{Dietsche1979}%
  \BibitemOpen
  \bibfield  {author} {\bibinfo {author} {\bibfnamefont {W.}~\bibnamefont {Dietsche}}\ and\ \bibinfo {author} {\bibfnamefont {H.}~\bibnamefont {Kinder}},\ }\bibfield  {title} {\bibinfo {title} {Spectroscopy of phonon scattering in glass},\ }\href {https://doi.org/10.1103/PhysRevLett.43.1413} {\bibfield  {journal} {\bibinfo  {journal} {Phys. Rev. Lett.}\ }\textbf {\bibinfo {volume} {43}},\ \bibinfo {pages} {1413} (\bibinfo {year} {1979})}\BibitemShut {NoStop}%
\bibitem [{\citenamefont {Koch}\ \emph {et~al.}(2023)\citenamefont {Koch}, \citenamefont {Mittleman}, \citenamefont {Ornik},\ and\ \citenamefont {Castro-Camus}}]{Koch2023}%
  \BibitemOpen
  \bibfield  {author} {\bibinfo {author} {\bibfnamefont {M.}~\bibnamefont {Koch}}, \bibinfo {author} {\bibfnamefont {D.~M.}\ \bibnamefont {Mittleman}}, \bibinfo {author} {\bibfnamefont {J.}~\bibnamefont {Ornik}},\ and\ \bibinfo {author} {\bibfnamefont {E.}~\bibnamefont {Castro-Camus}},\ }\bibfield  {title} {\bibinfo {title} {Terahertz time-domain spectroscopy},\ }\href {https://doi.org/10.1038/s43586-023-00232-z} {\bibfield  {journal} {\bibinfo  {journal} {Nature Reviews Methods Primers}\ }\textbf {\bibinfo {volume} {3}},\ \bibinfo {pages} {48} (\bibinfo {year} {2023})}\BibitemShut {NoStop}%
\bibitem [{\citenamefont {Chichibu}\ \emph {et~al.}(1998)\citenamefont {Chichibu}, \citenamefont {Abare}, \citenamefont {Minsky}, \citenamefont {Keller}, \citenamefont {Fleischer}, \citenamefont {Bowers}, \citenamefont {Hu}, \citenamefont {Mishra}, \citenamefont {Coldren}, \citenamefont {DenBaars},\ and\ \citenamefont {Sota}}]{Chichibu1998}%
  \BibitemOpen
  \bibfield  {author} {\bibinfo {author} {\bibfnamefont {S.~F.}\ \bibnamefont {Chichibu}}, \bibinfo {author} {\bibfnamefont {A.~C.}\ \bibnamefont {Abare}}, \bibinfo {author} {\bibfnamefont {M.~S.}\ \bibnamefont {Minsky}}, \bibinfo {author} {\bibfnamefont {S.}~\bibnamefont {Keller}}, \bibinfo {author} {\bibfnamefont {S.~B.}\ \bibnamefont {Fleischer}}, \bibinfo {author} {\bibfnamefont {J.~E.}\ \bibnamefont {Bowers}}, \bibinfo {author} {\bibfnamefont {E.}~\bibnamefont {Hu}}, \bibinfo {author} {\bibfnamefont {U.~K.}\ \bibnamefont {Mishra}}, \bibinfo {author} {\bibfnamefont {L.~A.}\ \bibnamefont {Coldren}}, \bibinfo {author} {\bibfnamefont {S.~P.}\ \bibnamefont {DenBaars}},\ and\ \bibinfo {author} {\bibfnamefont {T.}~\bibnamefont {Sota}},\ }\bibfield  {title} {\bibinfo {title} {Effective band gap inhomogeneity and piezoelectric field in ingan/gan multiquantum well structures},\ }\href {https://doi.org/10.1063/1.122350} {\bibfield  {journal} {\bibinfo  {journal} {Applied Physics Letters}\ }\textbf {\bibinfo
  {volume} {73}},\ \bibinfo {pages} {2006} (\bibinfo {year} {1998})},\ \Eprint {https://arxiv.org/abs/https://doi.org/10.1063/1.122350} {https://doi.org/10.1063/1.122350} \BibitemShut {NoStop}%
\bibitem [{\citenamefont {Wen}\ \emph {et~al.}(2011)\citenamefont {Wen}, \citenamefont {Guol}, \citenamefont {Chen}, \citenamefont {Sheu},\ and\ \citenamefont {Sun}}]{Wen2011}%
  \BibitemOpen
  \bibfield  {author} {\bibinfo {author} {\bibfnamefont {Y.-C.}\ \bibnamefont {Wen}}, \bibinfo {author} {\bibfnamefont {S.-H.}\ \bibnamefont {Guol}}, \bibinfo {author} {\bibfnamefont {H.-P.}\ \bibnamefont {Chen}}, \bibinfo {author} {\bibfnamefont {J.-K.}\ \bibnamefont {Sheu}},\ and\ \bibinfo {author} {\bibfnamefont {C.-K.}\ \bibnamefont {Sun}},\ }\bibfield  {title} {\bibinfo {title} {{Femtosecond ultrasonic spectroscopy using a piezoelectric nanolayer: Hypersound attenuation in vitreous silica films}},\ }\href {https://doi.org/10.1063/1.3620879} {\bibfield  {journal} {\bibinfo  {journal} {Applied Physics Letters}\ }\textbf {\bibinfo {volume} {99}},\ \bibinfo {pages} {051913} (\bibinfo {year} {2011})}\BibitemShut {NoStop}%
\bibitem [{\citenamefont {Lin}\ \emph {et~al.}(2007)\citenamefont {Lin}, \citenamefont {Lai}, \citenamefont {Pan}, \citenamefont {Chyi}, \citenamefont {Shi}, \citenamefont {Sun}, \citenamefont {Chang},\ and\ \citenamefont {Sun}}]{Lin2007}%
  \BibitemOpen
  \bibfield  {author} {\bibinfo {author} {\bibfnamefont {K.-H.}\ \bibnamefont {Lin}}, \bibinfo {author} {\bibfnamefont {C.-M.}\ \bibnamefont {Lai}}, \bibinfo {author} {\bibfnamefont {C.-C.}\ \bibnamefont {Pan}}, \bibinfo {author} {\bibfnamefont {J.-I.}\ \bibnamefont {Chyi}}, \bibinfo {author} {\bibfnamefont {J.-W.}\ \bibnamefont {Shi}}, \bibinfo {author} {\bibfnamefont {S.-Z.}\ \bibnamefont {Sun}}, \bibinfo {author} {\bibfnamefont {C.-F.}\ \bibnamefont {Chang}},\ and\ \bibinfo {author} {\bibfnamefont {C.-K.}\ \bibnamefont {Sun}},\ }\bibfield  {title} {\bibinfo {title} {Spatial manipulation of nanoacoustic waves with nanoscale spot sizes},\ }\href {https://doi.org/10.1038/nnano.2007.319} {\bibfield  {journal} {\bibinfo  {journal} {Nature nanotechnology}\ }\textbf {\bibinfo {volume} {2}},\ \bibinfo {pages} {704} (\bibinfo {year} {2007})}\BibitemShut {NoStop}%
\bibitem [{\citenamefont {Chern}\ \emph {et~al.}(2004{\natexlab{a}})\citenamefont {Chern}, \citenamefont {Sun}, \citenamefont {Sanders},\ and\ \citenamefont {Stanton}}]{Chern2004generation}%
  \BibitemOpen
  \bibfield  {author} {\bibinfo {author} {\bibfnamefont {G.-W.}\ \bibnamefont {Chern}}, \bibinfo {author} {\bibfnamefont {C.-K.}\ \bibnamefont {Sun}}, \bibinfo {author} {\bibfnamefont {G.~D.}\ \bibnamefont {Sanders}},\ and\ \bibinfo {author} {\bibfnamefont {C.~J.}\ \bibnamefont {Stanton}},\ }\bibfield  {title} {\bibinfo {title} {Generation of coherent acoustic phonons in nitride-based semiconductor nanostructures},\ }in\ \href@noop {} {\emph {\bibinfo {booktitle} {Ultrafast Dynamical Processes in Semiconductors}}}\ (\bibinfo  {publisher} {Springer},\ \bibinfo {year} {2004})\ pp.\ \bibinfo {pages} {339--394}\BibitemShut {NoStop}%
\bibitem [{\citenamefont {Malitson}(1965)}]{Malitson65}%
  \BibitemOpen
  \bibfield  {author} {\bibinfo {author} {\bibfnamefont {I.~H.}\ \bibnamefont {Malitson}},\ }\bibfield  {title} {\bibinfo {title} {Interspecimen comparison of the refractive index of fused silica},\ }\href {https://doi.org/10.1364/JOSA.55.001205} {\bibfield  {journal} {\bibinfo  {journal} {J. Opt. Soc. Am.}\ }\textbf {\bibinfo {volume} {55}},\ \bibinfo {pages} {1205} (\bibinfo {year} {1965})}\BibitemShut {NoStop}%
\bibitem [{\citenamefont {Mante}\ \emph {et~al.}(2013)\citenamefont {Mante}, \citenamefont {Chen}, \citenamefont {Wen}, \citenamefont {Sheu},\ and\ \citenamefont {Sun}}]{Mante2013thermal}%
  \BibitemOpen
  \bibfield  {author} {\bibinfo {author} {\bibfnamefont {P.-A.}\ \bibnamefont {Mante}}, \bibinfo {author} {\bibfnamefont {C.-C.}\ \bibnamefont {Chen}}, \bibinfo {author} {\bibfnamefont {Y.-C.}\ \bibnamefont {Wen}}, \bibinfo {author} {\bibfnamefont {J.-K.}\ \bibnamefont {Sheu}},\ and\ \bibinfo {author} {\bibfnamefont {C.-K.}\ \bibnamefont {Sun}},\ }\bibfield  {title} {\bibinfo {title} {Thermal boundary resistance between gan and cubic ice and thz acoustic attenuation spectrum of cubic ice from complex acoustic impedance measurements},\ }\href@noop {} {\bibfield  {journal} {\bibinfo  {journal} {Physical review letters}\ }\textbf {\bibinfo {volume} {111}},\ \bibinfo {pages} {225901} (\bibinfo {year} {2013})}\BibitemShut {NoStop}%
\bibitem [{\citenamefont {Chou}\ \emph {et~al.}(2019)\citenamefont {Chou}, \citenamefont {Lindsay}, \citenamefont {Maznev}, \citenamefont {Gandhi}, \citenamefont {Stokes}, \citenamefont {Forrest}, \citenamefont {Bensaoula}, \citenamefont {Nelson},\ and\ \citenamefont {Sun}}]{Chou2019}%
  \BibitemOpen
  \bibfield  {author} {\bibinfo {author} {\bibfnamefont {T.-H.}\ \bibnamefont {Chou}}, \bibinfo {author} {\bibfnamefont {L.}~\bibnamefont {Lindsay}}, \bibinfo {author} {\bibfnamefont {A.~A.}\ \bibnamefont {Maznev}}, \bibinfo {author} {\bibfnamefont {J.~S.}\ \bibnamefont {Gandhi}}, \bibinfo {author} {\bibfnamefont {D.~W.}\ \bibnamefont {Stokes}}, \bibinfo {author} {\bibfnamefont {R.~L.}\ \bibnamefont {Forrest}}, \bibinfo {author} {\bibfnamefont {A.}~\bibnamefont {Bensaoula}}, \bibinfo {author} {\bibfnamefont {K.~A.}\ \bibnamefont {Nelson}},\ and\ \bibinfo {author} {\bibfnamefont {C.-K.}\ \bibnamefont {Sun}},\ }\bibfield  {title} {\bibinfo {title} {Long mean free paths of room-temperature thz acoustic phonons in a high thermal conductivity material},\ }\href {https://doi.org/10.1103/PhysRevB.100.094302} {\bibfield  {journal} {\bibinfo  {journal} {Phys. Rev. B}\ }\textbf {\bibinfo {volume} {100}},\ \bibinfo {pages} {094302} (\bibinfo {year} {2019})}\BibitemShut {NoStop}%
\bibitem [{\citenamefont {Mante}\ \emph {et~al.}(2015)\citenamefont {Mante}, \citenamefont {Huang}, \citenamefont {Yang}, \citenamefont {Liu}, \citenamefont {Maznev}, \citenamefont {Sheu},\ and\ \citenamefont {Sun}}]{Mante2015}%
  \BibitemOpen
  \bibfield  {author} {\bibinfo {author} {\bibfnamefont {P.-A.}\ \bibnamefont {Mante}}, \bibinfo {author} {\bibfnamefont {Y.-R.}\ \bibnamefont {Huang}}, \bibinfo {author} {\bibfnamefont {S.-C.}\ \bibnamefont {Yang}}, \bibinfo {author} {\bibfnamefont {T.-M.}\ \bibnamefont {Liu}}, \bibinfo {author} {\bibfnamefont {A.~A.}\ \bibnamefont {Maznev}}, \bibinfo {author} {\bibfnamefont {J.-K.}\ \bibnamefont {Sheu}},\ and\ \bibinfo {author} {\bibfnamefont {C.-K.}\ \bibnamefont {Sun}},\ }\bibfield  {title} {\bibinfo {title} {Thz acoustic phonon spectroscopy and nanoscopy by using piezoelectric semiconductor heterostructures},\ }\href {https://doi.org/https://doi.org/10.1016/j.ultras.2014.09.020} {\bibfield  {journal} {\bibinfo  {journal} {Ultrasonics}\ }\textbf {\bibinfo {volume} {56}},\ \bibinfo {pages} {52} (\bibinfo {year} {2015})}\BibitemShut {NoStop}%
\bibitem [{\citenamefont {Sun}\ \emph {et~al.}(2000)\citenamefont {Sun}, \citenamefont {Liang},\ and\ \citenamefont {Yu}}]{Sun2000}%
  \BibitemOpen
  \bibfield  {author} {\bibinfo {author} {\bibfnamefont {C.-K.}\ \bibnamefont {Sun}}, \bibinfo {author} {\bibfnamefont {J.-C.}\ \bibnamefont {Liang}},\ and\ \bibinfo {author} {\bibfnamefont {X.-Y.}\ \bibnamefont {Yu}},\ }\bibfield  {title} {\bibinfo {title} {Coherent acoustic phonon oscillations in semiconductor multiple quantum wells with piezoelectric fields},\ }\href {https://doi.org/10.1103/PhysRevLett.84.179} {\bibfield  {journal} {\bibinfo  {journal} {Phys. Rev. Lett.}\ }\textbf {\bibinfo {volume} {84}},\ \bibinfo {pages} {179} (\bibinfo {year} {2000})}\BibitemShut {NoStop}%
\bibitem [{\citenamefont {Chern}\ \emph {et~al.}(2004{\natexlab{b}})\citenamefont {Chern}, \citenamefont {Lin},\ and\ \citenamefont {Sun}}]{Chern2004}%
  \BibitemOpen
  \bibfield  {author} {\bibinfo {author} {\bibfnamefont {G.-W.}\ \bibnamefont {Chern}}, \bibinfo {author} {\bibfnamefont {K.-H.}\ \bibnamefont {Lin}},\ and\ \bibinfo {author} {\bibfnamefont {C.-K.}\ \bibnamefont {Sun}},\ }\bibfield  {title} {\bibinfo {title} {Transmission of light through quantum heterostructures modulated by coherent acoustic phonons},\ }\href {https://doi.org/10.1063/1.1637957} {\bibfield  {journal} {\bibinfo  {journal} {Journal of Applied Physics}\ }\textbf {\bibinfo {volume} {95}},\ \bibinfo {pages} {1114} (\bibinfo {year} {2004}{\natexlab{b}})},\ \Eprint {https://arxiv.org/abs/https://doi.org/10.1063/1.1637957} {https://doi.org/10.1063/1.1637957} \BibitemShut {NoStop}%
\bibitem [{\citenamefont {{Kung-Hsuan Lin}}\ \emph {et~al.}(2005)\citenamefont {{Kung-Hsuan Lin}}, \citenamefont {{Gia-Wei Chern}}, \citenamefont {{Cheng-Ta Yu}}, \citenamefont {{Tzu-Ming Liu}}, \citenamefont {{Chang-Chi Pan}}, \citenamefont {{Guan-Ting Chen}}, \citenamefont {{Jen-Inn Chyi}}, \citenamefont {{Sheng-Wen Huang}}, \citenamefont {{Pai-Chi Li}},\ and\ \citenamefont {{Chi-Kuang Sun}}}]{Lin2005}%
  \BibitemOpen
  \bibfield  {author} {\bibinfo {author} {\bibnamefont {{Kung-Hsuan Lin}}}, \bibinfo {author} {\bibnamefont {{Gia-Wei Chern}}}, \bibinfo {author} {\bibnamefont {{Cheng-Ta Yu}}}, \bibinfo {author} {\bibnamefont {{Tzu-Ming Liu}}}, \bibinfo {author} {\bibnamefont {{Chang-Chi Pan}}}, \bibinfo {author} {\bibnamefont {{Guan-Ting Chen}}}, \bibinfo {author} {\bibnamefont {{Jen-Inn Chyi}}}, \bibinfo {author} {\bibnamefont {{Sheng-Wen Huang}}}, \bibinfo {author} {\bibnamefont {{Pai-Chi Li}}},\ and\ \bibinfo {author} {\bibnamefont {{Chi-Kuang Sun}}},\ }\bibfield  {title} {\bibinfo {title} {Optical piezoelectric transducer for nano-ultrasonics},\ }\href {https://doi.org/10.1109/TUFFC.2005.1509800} {\bibfield  {journal} {\bibinfo  {journal} {IEEE Transactions on Ultrasonics, Ferroelectrics, and Frequency Control}\ }\textbf {\bibinfo {volume} {52}},\ \bibinfo {pages} {1404} (\bibinfo {year} {2005})}\BibitemShut {NoStop}%
\bibitem [{\citenamefont {Vacher}\ \emph {et~al.}(2006)\citenamefont {Vacher}, \citenamefont {Ayrinhac}, \citenamefont {Foret}, \citenamefont {Ruffl\'e},\ and\ \citenamefont {Courtens}}]{Vacher2006}%
  \BibitemOpen
  \bibfield  {author} {\bibinfo {author} {\bibfnamefont {R.}~\bibnamefont {Vacher}}, \bibinfo {author} {\bibfnamefont {S.}~\bibnamefont {Ayrinhac}}, \bibinfo {author} {\bibfnamefont {M.}~\bibnamefont {Foret}}, \bibinfo {author} {\bibfnamefont {B.}~\bibnamefont {Ruffl\'e}},\ and\ \bibinfo {author} {\bibfnamefont {E.}~\bibnamefont {Courtens}},\ }\bibfield  {title} {\bibinfo {title} {Finite size effects in brillouin scattering from silica glass},\ }\href {https://doi.org/10.1103/PhysRevB.74.012203} {\bibfield  {journal} {\bibinfo  {journal} {Phys. Rev. B}\ }\textbf {\bibinfo {volume} {74}},\ \bibinfo {pages} {012203} (\bibinfo {year} {2006})}\BibitemShut {NoStop}%
\bibitem [{\citenamefont {Masciovecchio}\ \emph {et~al.}(2004)\citenamefont {Masciovecchio}, \citenamefont {Gessini}, \citenamefont {Di~Fonzo}, \citenamefont {Comez}, \citenamefont {Santucci},\ and\ \citenamefont {Fioretto}}]{Masciovecchio2004}%
  \BibitemOpen
  \bibfield  {author} {\bibinfo {author} {\bibfnamefont {C.}~\bibnamefont {Masciovecchio}}, \bibinfo {author} {\bibfnamefont {A.}~\bibnamefont {Gessini}}, \bibinfo {author} {\bibfnamefont {S.}~\bibnamefont {Di~Fonzo}}, \bibinfo {author} {\bibfnamefont {L.}~\bibnamefont {Comez}}, \bibinfo {author} {\bibfnamefont {S.~C.}\ \bibnamefont {Santucci}},\ and\ \bibinfo {author} {\bibfnamefont {D.}~\bibnamefont {Fioretto}},\ }\bibfield  {title} {\bibinfo {title} {Inelastic ultraviolet scattering from high frequency acoustic modes in glasses},\ }\href {https://doi.org/10.1103/PhysRevLett.92.247401} {\bibfield  {journal} {\bibinfo  {journal} {Phys. Rev. Lett.}\ }\textbf {\bibinfo {volume} {92}},\ \bibinfo {pages} {247401} (\bibinfo {year} {2004})}\BibitemShut {NoStop}%
\bibitem [{\citenamefont {Benassi}\ \emph {et~al.}(2005)\citenamefont {Benassi}, \citenamefont {Caponi}, \citenamefont {Eramo}, \citenamefont {Fontana}, \citenamefont {Giugni}, \citenamefont {Nardone}, \citenamefont {Sampoli},\ and\ \citenamefont {Viliani}}]{Benassi2005}%
  \BibitemOpen
  \bibfield  {author} {\bibinfo {author} {\bibfnamefont {P.}~\bibnamefont {Benassi}}, \bibinfo {author} {\bibfnamefont {S.}~\bibnamefont {Caponi}}, \bibinfo {author} {\bibfnamefont {R.}~\bibnamefont {Eramo}}, \bibinfo {author} {\bibfnamefont {A.}~\bibnamefont {Fontana}}, \bibinfo {author} {\bibfnamefont {A.}~\bibnamefont {Giugni}}, \bibinfo {author} {\bibfnamefont {M.}~\bibnamefont {Nardone}}, \bibinfo {author} {\bibfnamefont {M.}~\bibnamefont {Sampoli}},\ and\ \bibinfo {author} {\bibfnamefont {G.}~\bibnamefont {Viliani}},\ }\bibfield  {title} {\bibinfo {title} {Sound attenuation in a unexplored frequency region: Brillouin ultraviolet light scattering measurements in $v\text{\ensuremath{-}}\mathrm{Si}{\mathrm{o}}_{2}$},\ }\href {https://doi.org/10.1103/PhysRevB.71.172201} {\bibfield  {journal} {\bibinfo  {journal} {Phys. Rev. B}\ }\textbf {\bibinfo {volume} {71}},\ \bibinfo {pages} {172201} (\bibinfo {year} {2005})}\BibitemShut {NoStop}%
\bibitem [{\citenamefont {Ruffl\'e}\ \emph {et~al.}(2011)\citenamefont {Ruffl\'e}, \citenamefont {Courtens},\ and\ \citenamefont {Foret}}]{Ruffle2011}%
  \BibitemOpen
  \bibfield  {author} {\bibinfo {author} {\bibfnamefont {B.}~\bibnamefont {Ruffl\'e}}, \bibinfo {author} {\bibfnamefont {E.}~\bibnamefont {Courtens}},\ and\ \bibinfo {author} {\bibfnamefont {M.}~\bibnamefont {Foret}},\ }\bibfield  {title} {\bibinfo {title} {Inelastic ultraviolet brillouin scattering from superpolished vitreous silica},\ }\href {https://doi.org/10.1103/PhysRevB.84.132201} {\bibfield  {journal} {\bibinfo  {journal} {Phys. Rev. B}\ }\textbf {\bibinfo {volume} {84}},\ \bibinfo {pages} {132201} (\bibinfo {year} {2011})}\BibitemShut {NoStop}%
\bibitem [{\citenamefont {Masciovecchio}\ \emph {et~al.}(2006)\citenamefont {Masciovecchio}, \citenamefont {Baldi}, \citenamefont {Caponi}, \citenamefont {Comez}, \citenamefont {Di~Fonzo}, \citenamefont {Fioretto}, \citenamefont {Fontana}, \citenamefont {Gessini}, \citenamefont {Santucci}, \citenamefont {Sette}, \citenamefont {Viliani}, \citenamefont {Vilmercati},\ and\ \citenamefont {Ruocco}}]{Masciovecchio2006}%
  \BibitemOpen
  \bibfield  {author} {\bibinfo {author} {\bibfnamefont {C.}~\bibnamefont {Masciovecchio}}, \bibinfo {author} {\bibfnamefont {G.}~\bibnamefont {Baldi}}, \bibinfo {author} {\bibfnamefont {S.}~\bibnamefont {Caponi}}, \bibinfo {author} {\bibfnamefont {L.}~\bibnamefont {Comez}}, \bibinfo {author} {\bibfnamefont {S.}~\bibnamefont {Di~Fonzo}}, \bibinfo {author} {\bibfnamefont {D.}~\bibnamefont {Fioretto}}, \bibinfo {author} {\bibfnamefont {A.}~\bibnamefont {Fontana}}, \bibinfo {author} {\bibfnamefont {A.}~\bibnamefont {Gessini}}, \bibinfo {author} {\bibfnamefont {S.~C.}\ \bibnamefont {Santucci}}, \bibinfo {author} {\bibfnamefont {F.}~\bibnamefont {Sette}}, \bibinfo {author} {\bibfnamefont {G.}~\bibnamefont {Viliani}}, \bibinfo {author} {\bibfnamefont {P.}~\bibnamefont {Vilmercati}},\ and\ \bibinfo {author} {\bibfnamefont {G.}~\bibnamefont {Ruocco}},\ }\bibfield  {title} {\bibinfo {title} {Evidence for a crossover in the frequency dependence of the acoustic attenuation in vitreous silica},\ }\href
  {https://doi.org/10.1103/PhysRevLett.97.035501} {\bibfield  {journal} {\bibinfo  {journal} {Phys. Rev. Lett.}\ }\textbf {\bibinfo {volume} {97}},\ \bibinfo {pages} {035501} (\bibinfo {year} {2006})}\BibitemShut {NoStop}%
\bibitem [{\citenamefont {Ayrinhac}\ \emph {et~al.}(2011)\citenamefont {Ayrinhac}, \citenamefont {Foret}, \citenamefont {Devos}, \citenamefont {Ruffl\'e}, \citenamefont {Courtens},\ and\ \citenamefont {Vacher}}]{Ayrinhac2011a}%
  \BibitemOpen
  \bibfield  {author} {\bibinfo {author} {\bibfnamefont {S.}~\bibnamefont {Ayrinhac}}, \bibinfo {author} {\bibfnamefont {M.}~\bibnamefont {Foret}}, \bibinfo {author} {\bibfnamefont {A.}~\bibnamefont {Devos}}, \bibinfo {author} {\bibfnamefont {B.}~\bibnamefont {Ruffl\'e}}, \bibinfo {author} {\bibfnamefont {E.}~\bibnamefont {Courtens}},\ and\ \bibinfo {author} {\bibfnamefont {R.}~\bibnamefont {Vacher}},\ }\bibfield  {title} {\bibinfo {title} {Subterahertz hypersound attenuation in silica glass studied via picosecond acoustics},\ }\href {https://doi.org/10.1103/PhysRevB.83.014204} {\bibfield  {journal} {\bibinfo  {journal} {Phys. Rev. B}\ }\textbf {\bibinfo {volume} {83}},\ \bibinfo {pages} {014204} (\bibinfo {year} {2011})}\BibitemShut {NoStop}%
\bibitem [{\citenamefont {Baldi}\ \emph {et~al.}(2011)\citenamefont {Baldi}, \citenamefont {Giordano},\ and\ \citenamefont {Monaco}}]{Baldi2011}%
  \BibitemOpen
  \bibfield  {author} {\bibinfo {author} {\bibfnamefont {G.}~\bibnamefont {Baldi}}, \bibinfo {author} {\bibfnamefont {V.~M.}\ \bibnamefont {Giordano}},\ and\ \bibinfo {author} {\bibfnamefont {G.}~\bibnamefont {Monaco}},\ }\bibfield  {title} {\bibinfo {title} {Elastic anomalies at terahertz frequencies and excess density of vibrational states in silica glass},\ }\href {https://doi.org/10.1103/PhysRevB.83.174203} {\bibfield  {journal} {\bibinfo  {journal} {Phys. Rev. B}\ }\textbf {\bibinfo {volume} {83}},\ \bibinfo {pages} {174203} (\bibinfo {year} {2011})}\BibitemShut {NoStop}%
\bibitem [{\citenamefont {Zhu}\ \emph {et~al.}(1991)\citenamefont {Zhu}, \citenamefont {Maris},\ and\ \citenamefont {Tauc}}]{Zhu1991}%
  \BibitemOpen
  \bibfield  {author} {\bibinfo {author} {\bibfnamefont {T.~C.}\ \bibnamefont {Zhu}}, \bibinfo {author} {\bibfnamefont {H.~J.}\ \bibnamefont {Maris}},\ and\ \bibinfo {author} {\bibfnamefont {J.}~\bibnamefont {Tauc}},\ }\bibfield  {title} {\bibinfo {title} {Attenuation of longitudinal-acoustic phonons in amorphous ${\mathrm{sio}}_{2}$ at frequencies up to 440 ghz},\ }\href {https://doi.org/10.1103/PhysRevB.44.4281} {\bibfield  {journal} {\bibinfo  {journal} {Phys. Rev. B}\ }\textbf {\bibinfo {volume} {44}},\ \bibinfo {pages} {4281} (\bibinfo {year} {1991})}\BibitemShut {NoStop}%
\bibitem [{\citenamefont {Foret}\ \emph {et~al.}(1996)\citenamefont {Foret}, \citenamefont {Courtens}, \citenamefont {Vacher},\ and\ \citenamefont {Suck}}]{Foret1996}%
  \BibitemOpen
  \bibfield  {author} {\bibinfo {author} {\bibfnamefont {M.}~\bibnamefont {Foret}}, \bibinfo {author} {\bibfnamefont {E.}~\bibnamefont {Courtens}}, \bibinfo {author} {\bibfnamefont {R.}~\bibnamefont {Vacher}},\ and\ \bibinfo {author} {\bibfnamefont {J.-B.}\ \bibnamefont {Suck}},\ }\bibfield  {title} {\bibinfo {title} {Scattering investigation of acoustic localization in fused silica},\ }\href {https://doi.org/10.1103/PhysRevLett.77.3831} {\bibfield  {journal} {\bibinfo  {journal} {Phys. Rev. Lett.}\ }\textbf {\bibinfo {volume} {77}},\ \bibinfo {pages} {3831} (\bibinfo {year} {1996})}\BibitemShut {NoStop}%
\bibitem [{\citenamefont {Monaco}\ and\ \citenamefont {Mossa}(2009)}]{Monaco2009b}%
  \BibitemOpen
  \bibfield  {author} {\bibinfo {author} {\bibfnamefont {G.}~\bibnamefont {Monaco}}\ and\ \bibinfo {author} {\bibfnamefont {S.}~\bibnamefont {Mossa}},\ }\bibfield  {title} {\bibinfo {title} {Anomalous properties of the acoustic excitations in glasses on the mesoscopic length scale},\ }\href {https://doi.org/10.1073/pnas.0903922106} {\bibfield  {journal} {\bibinfo  {journal} {Proceedings of the National Academy of Sciences}\ }\textbf {\bibinfo {volume} {106}},\ \bibinfo {pages} {16907} (\bibinfo {year} {2009})},\ \Eprint {https://arxiv.org/abs/https://www.pnas.org/doi/pdf/10.1073/pnas.0903922106} {https://www.pnas.org/doi/pdf/10.1073/pnas.0903922106} \BibitemShut {NoStop}%
\bibitem [{\citenamefont {Buchenau}\ \emph {et~al.}(1992)\citenamefont {Buchenau}, \citenamefont {Galperin}, \citenamefont {Gurevich}, \citenamefont {Parshin}, \citenamefont {Ramos},\ and\ \citenamefont {Schober}}]{Buchenau1992}%
  \BibitemOpen
  \bibfield  {author} {\bibinfo {author} {\bibfnamefont {U.}~\bibnamefont {Buchenau}}, \bibinfo {author} {\bibfnamefont {Y.~M.}\ \bibnamefont {Galperin}}, \bibinfo {author} {\bibfnamefont {V.~L.}\ \bibnamefont {Gurevich}}, \bibinfo {author} {\bibfnamefont {D.~A.}\ \bibnamefont {Parshin}}, \bibinfo {author} {\bibfnamefont {M.~A.}\ \bibnamefont {Ramos}},\ and\ \bibinfo {author} {\bibfnamefont {H.~R.}\ \bibnamefont {Schober}},\ }\bibfield  {title} {\bibinfo {title} {Interaction of soft modes and sound waves in glasses},\ }\href {https://doi.org/10.1103/PhysRevB.46.2798} {\bibfield  {journal} {\bibinfo  {journal} {Phys. Rev. B}\ }\textbf {\bibinfo {volume} {46}},\ \bibinfo {pages} {2798} (\bibinfo {year} {1992})}\BibitemShut {NoStop}%
\bibitem [{\citenamefont {Schober}(2011)}]{Schober2011}%
  \BibitemOpen
  \bibfield  {author} {\bibinfo {author} {\bibfnamefont {H.~R.}\ \bibnamefont {Schober}},\ }\bibfield  {title} {\bibinfo {title} {Quasi-localized vibrations and phonon damping in glasses},\ }\href {https://doi.org/https://doi.org/10.1016/j.jnoncrysol.2010.07.036} {\bibfield  {journal} {\bibinfo  {journal} {Journal of Non-Crystalline Solids}\ }\textbf {\bibinfo {volume} {357}},\ \bibinfo {pages} {501 } (\bibinfo {year} {2011})},\ \bibinfo {note} {6th International Discussion Meeting on Relaxation in Complex Systems}\BibitemShut {NoStop}%
\bibitem [{\citenamefont {Mizuno}\ \emph {et~al.}(2017)\citenamefont {Mizuno}, \citenamefont {Shiba},\ and\ \citenamefont {Ikeda}}]{Mizuno2017}%
  \BibitemOpen
  \bibfield  {author} {\bibinfo {author} {\bibfnamefont {H.}~\bibnamefont {Mizuno}}, \bibinfo {author} {\bibfnamefont {H.}~\bibnamefont {Shiba}},\ and\ \bibinfo {author} {\bibfnamefont {A.}~\bibnamefont {Ikeda}},\ }\bibfield  {title} {\bibinfo {title} {Continuum limit of the vibrational properties of amorphous solids},\ }\href {https://doi.org/10.1073/pnas.1709015114} {\bibfield  {journal} {\bibinfo  {journal} {Proceedings of the National Academy of Sciences}\ }\textbf {\bibinfo {volume} {114}},\ \bibinfo {pages} {E9767} (\bibinfo {year} {2017})},\ \Eprint {https://arxiv.org/abs/https://www.pnas.org/doi/pdf/10.1073/pnas.1709015114} {https://www.pnas.org/doi/pdf/10.1073/pnas.1709015114} \BibitemShut {NoStop}%
\bibitem [{\citenamefont {Lerner}\ and\ \citenamefont {Bouchbinder}(2021)}]{Lerner2021}%
  \BibitemOpen
  \bibfield  {author} {\bibinfo {author} {\bibfnamefont {E.}~\bibnamefont {Lerner}}\ and\ \bibinfo {author} {\bibfnamefont {E.}~\bibnamefont {Bouchbinder}},\ }\bibfield  {title} {\bibinfo {title} {Low-energy quasilocalized excitations in structural glasses},\ }\href {https://doi.org/10.1063/5.0069477} {\bibfield  {journal} {\bibinfo  {journal} {The Journal of Chemical Physics}\ }\textbf {\bibinfo {volume} {155}},\ \bibinfo {pages} {200901} (\bibinfo {year} {2021})},\ \Eprint {https://arxiv.org/abs/https://doi.org/10.1063/5.0069477} {https://doi.org/10.1063/5.0069477} \BibitemShut {NoStop}%
\bibitem [{\citenamefont {Richard}\ \emph {et~al.}(2020)\citenamefont {Richard}, \citenamefont {Gonz\'alez-L\'opez}, \citenamefont {Kapteijns}, \citenamefont {Pater}, \citenamefont {Vaknin}, \citenamefont {Bouchbinder},\ and\ \citenamefont {Lerner}}]{Richard2020}%
  \BibitemOpen
  \bibfield  {author} {\bibinfo {author} {\bibfnamefont {D.}~\bibnamefont {Richard}}, \bibinfo {author} {\bibfnamefont {K.}~\bibnamefont {Gonz\'alez-L\'opez}}, \bibinfo {author} {\bibfnamefont {G.}~\bibnamefont {Kapteijns}}, \bibinfo {author} {\bibfnamefont {R.}~\bibnamefont {Pater}}, \bibinfo {author} {\bibfnamefont {T.}~\bibnamefont {Vaknin}}, \bibinfo {author} {\bibfnamefont {E.}~\bibnamefont {Bouchbinder}},\ and\ \bibinfo {author} {\bibfnamefont {E.}~\bibnamefont {Lerner}},\ }\bibfield  {title} {\bibinfo {title} {Universality of the nonphononic vibrational spectrum across different classes of computer glasses},\ }\href {https://doi.org/10.1103/PhysRevLett.125.085502} {\bibfield  {journal} {\bibinfo  {journal} {Phys. Rev. Lett.}\ }\textbf {\bibinfo {volume} {125}},\ \bibinfo {pages} {085502} (\bibinfo {year} {2020})}\BibitemShut {NoStop}%
\bibitem [{\citenamefont {Buchenau}\ \emph {et~al.}(1991)\citenamefont {Buchenau}, \citenamefont {Galperin}, \citenamefont {Gurevich},\ and\ \citenamefont {Schober}}]{Buchenau1991}%
  \BibitemOpen
  \bibfield  {author} {\bibinfo {author} {\bibfnamefont {U.}~\bibnamefont {Buchenau}}, \bibinfo {author} {\bibfnamefont {Y.~M.}\ \bibnamefont {Galperin}}, \bibinfo {author} {\bibfnamefont {V.~L.}\ \bibnamefont {Gurevich}},\ and\ \bibinfo {author} {\bibfnamefont {H.~R.}\ \bibnamefont {Schober}},\ }\bibfield  {title} {\bibinfo {title} {Anharmonic potentials and vibrational localization in glasses},\ }\href {https://doi.org/10.1103/PhysRevB.43.5039} {\bibfield  {journal} {\bibinfo  {journal} {Phys. Rev. B}\ }\textbf {\bibinfo {volume} {43}},\ \bibinfo {pages} {5039} (\bibinfo {year} {1991})}\BibitemShut {NoStop}%
\bibitem [{\citenamefont {Gil}\ \emph {et~al.}(1993)\citenamefont {Gil}, \citenamefont {Ramos}, \citenamefont {Bringer},\ and\ \citenamefont {Buchenau}}]{Gil1993}%
  \BibitemOpen
  \bibfield  {author} {\bibinfo {author} {\bibfnamefont {L.}~\bibnamefont {Gil}}, \bibinfo {author} {\bibfnamefont {M.~A.}\ \bibnamefont {Ramos}}, \bibinfo {author} {\bibfnamefont {A.}~\bibnamefont {Bringer}},\ and\ \bibinfo {author} {\bibfnamefont {U.}~\bibnamefont {Buchenau}},\ }\bibfield  {title} {\bibinfo {title} {Low-temperature specific heat and thermal conductivity of glasses},\ }\href {https://doi.org/10.1103/PhysRevLett.70.182} {\bibfield  {journal} {\bibinfo  {journal} {Phys. Rev. Lett.}\ }\textbf {\bibinfo {volume} {70}},\ \bibinfo {pages} {182} (\bibinfo {year} {1993})}\BibitemShut {NoStop}%
\bibitem [{\citenamefont {Ruffl\'e}\ \emph {et~al.}(2008)\citenamefont {Ruffl\'e}, \citenamefont {Parshin}, \citenamefont {Courtens},\ and\ \citenamefont {Vacher}}]{Ruffle2008}%
  \BibitemOpen
  \bibfield  {author} {\bibinfo {author} {\bibfnamefont {B.}~\bibnamefont {Ruffl\'e}}, \bibinfo {author} {\bibfnamefont {D.~A.}\ \bibnamefont {Parshin}}, \bibinfo {author} {\bibfnamefont {E.}~\bibnamefont {Courtens}},\ and\ \bibinfo {author} {\bibfnamefont {R.}~\bibnamefont {Vacher}},\ }\bibfield  {title} {\bibinfo {title} {Boson peak and its relation to acoustic attenuation in glasses},\ }\href {https://doi.org/10.1103/PhysRevLett.100.015501} {\bibfield  {journal} {\bibinfo  {journal} {Phys. Rev. Lett.}\ }\textbf {\bibinfo {volume} {100}},\ \bibinfo {pages} {015501} (\bibinfo {year} {2008})}\BibitemShut {NoStop}%
\bibitem [{\citenamefont {Lerner}\ and\ \citenamefont {Bouchbinder}(2018)}]{Lerner2018}%
  \BibitemOpen
  \bibfield  {author} {\bibinfo {author} {\bibfnamefont {E.}~\bibnamefont {Lerner}}\ and\ \bibinfo {author} {\bibfnamefont {E.}~\bibnamefont {Bouchbinder}},\ }\bibfield  {title} {\bibinfo {title} {Frustration-induced internal stresses are responsible for quasilocalized modes in structural glasses},\ }\href {https://doi.org/10.1103/PhysRevE.97.032140} {\bibfield  {journal} {\bibinfo  {journal} {Phys. Rev. E}\ }\textbf {\bibinfo {volume} {97}},\ \bibinfo {pages} {032140} (\bibinfo {year} {2018})}\BibitemShut {NoStop}%
\bibitem [{\citenamefont {Kapteijns}\ \emph {et~al.}(2021)\citenamefont {Kapteijns}, \citenamefont {Bouchbinder},\ and\ \citenamefont {Lerner}}]{Kapteijns2021b}%
  \BibitemOpen
  \bibfield  {author} {\bibinfo {author} {\bibfnamefont {G.}~\bibnamefont {Kapteijns}}, \bibinfo {author} {\bibfnamefont {E.}~\bibnamefont {Bouchbinder}},\ and\ \bibinfo {author} {\bibfnamefont {E.}~\bibnamefont {Lerner}},\ }\bibfield  {title} {\bibinfo {title} {Unified quantifier of mechanical disorder in solids},\ }\href {https://doi.org/10.1103/PhysRevE.104.035001} {\bibfield  {journal} {\bibinfo  {journal} {Phys. Rev. E}\ }\textbf {\bibinfo {volume} {104}},\ \bibinfo {pages} {035001} (\bibinfo {year} {2021})}\BibitemShut {NoStop}%
\bibitem [{\citenamefont {Buchenau}\ \emph {et~al.}(1984)\citenamefont {Buchenau}, \citenamefont {N\"ucker},\ and\ \citenamefont {Dianoux}}]{Buchenau1984}%
  \BibitemOpen
  \bibfield  {author} {\bibinfo {author} {\bibfnamefont {U.}~\bibnamefont {Buchenau}}, \bibinfo {author} {\bibfnamefont {N.}~\bibnamefont {N\"ucker}},\ and\ \bibinfo {author} {\bibfnamefont {A.~J.}\ \bibnamefont {Dianoux}},\ }\bibfield  {title} {\bibinfo {title} {Neutron scattering study of the low-frequency vibrations in vitreous silica},\ }\href {https://doi.org/10.1103/PhysRevLett.53.2316} {\bibfield  {journal} {\bibinfo  {journal} {Phys. Rev. Lett.}\ }\textbf {\bibinfo {volume} {53}},\ \bibinfo {pages} {2316} (\bibinfo {year} {1984})}\BibitemShut {NoStop}%
\bibitem [{\citenamefont {Schirmacher}\ \emph {et~al.}(2007)\citenamefont {Schirmacher}, \citenamefont {Ruocco},\ and\ \citenamefont {Scopigno}}]{Schirmacher2007}%
  \BibitemOpen
  \bibfield  {author} {\bibinfo {author} {\bibfnamefont {W.}~\bibnamefont {Schirmacher}}, \bibinfo {author} {\bibfnamefont {G.}~\bibnamefont {Ruocco}},\ and\ \bibinfo {author} {\bibfnamefont {T.}~\bibnamefont {Scopigno}},\ }\bibfield  {title} {\bibinfo {title} {Acoustic attenuation in glasses and its relation with the boson peak},\ }\href {https://doi.org/10.1103/PhysRevLett.98.025501} {\bibfield  {journal} {\bibinfo  {journal} {Phys. Rev. Lett.}\ }\textbf {\bibinfo {volume} {98}},\ \bibinfo {pages} {025501} (\bibinfo {year} {2007})}\BibitemShut {NoStop}%
\bibitem [{\citenamefont {Buchenau}(2014)}]{Buchenau2014}%
  \BibitemOpen
  \bibfield  {author} {\bibinfo {author} {\bibfnamefont {U.}~\bibnamefont {Buchenau}},\ }\bibfield  {title} {\bibinfo {title} {Evaluation of x-ray brillouin scattering data},\ }\href {https://doi.org/10.1103/PhysRevE.90.062319} {\bibfield  {journal} {\bibinfo  {journal} {Phys. Rev. E}\ }\textbf {\bibinfo {volume} {90}},\ \bibinfo {pages} {062319} (\bibinfo {year} {2014})}\BibitemShut {NoStop}%
\bibitem [{\citenamefont {Baldi}\ \emph {et~al.}(2016)\citenamefont {Baldi}, \citenamefont {Giordano}, \citenamefont {Ruta},\ and\ \citenamefont {Monaco}}]{Baldi2016}%
  \BibitemOpen
  \bibfield  {author} {\bibinfo {author} {\bibfnamefont {G.}~\bibnamefont {Baldi}}, \bibinfo {author} {\bibfnamefont {V.~M.}\ \bibnamefont {Giordano}}, \bibinfo {author} {\bibfnamefont {B.}~\bibnamefont {Ruta}},\ and\ \bibinfo {author} {\bibfnamefont {G.}~\bibnamefont {Monaco}},\ }\bibfield  {title} {\bibinfo {title} {On the nontrivial wave-vector dependence of the elastic modulus of glasses},\ }\href {https://doi.org/10.1103/PhysRevB.93.144204} {\bibfield  {journal} {\bibinfo  {journal} {Phys. Rev. B}\ }\textbf {\bibinfo {volume} {93}},\ \bibinfo {pages} {144204} (\bibinfo {year} {2016})}\BibitemShut {NoStop}%
\end{thebibliography}
%

\end{document}